\documentclass[preprint2]{aastex}

\shorttitle{Wind Law for WR~136}
\shortauthors{Ignace, Quigley, Cassinelli}

\received{2001 August 17}

\begin{document}
\newcommand{\ul}{\underline} % \ul
\def \etal   {\hbox{\it et~al.\/}}
\def \ie   {\hbox{i.e.\/}}
\def \eg   {\hbox{e.g.\/}}
\def \kms    {~km~s$^{-1}$}
\def \Mdot     {{$M_{\odot}$}}
\def \msunyr   {{$M_{\odot}$~yr$^{-1}$}}
\def \bb       {\hbox{$\beta$}}
\def \lam      {\hbox{$\lambda$}}
\def \Llam      {\hbox{$L_\lambda$}}
\def \wlam      {\hbox{$w_\lambda$}}
\def \kaplam    {\hbox{$\kappa_\lambda$}}
\def \Tw      {\hbox{$T_w$}}
\def \Tstar      {\hbox{$T_*$}}
\def \gamav   {\hbox{$\gamma_{\rm v}$}}
\def \gamaF   {\hbox{$\gamma_F$}}
\def \Dclump   {\hbox{$D_c$}}
\def \dellam  {\hbox{$\delta \lambda$}}
\def \delw    {\hbox{$\delta w$}}
\def \flam    {\hbox{$f_\lambda$}}
\def \Flam    {\hbox{$F_\lambda$}}
\def \lamflam {\hbox{$\lambda f_\lambda$}}
\def \gammaf  {\hbox{$\gamma_F$}}
\def \gammaw  {\hbox{$\gamma_W$}}
\def \taulam    {\hbox{$\tau_\lambda$}}
\def \tauff    {\hbox{$\tau_{\rm ff}$}}
\def \tableref#1{Table~\ref{#1}}
\def \figureref#1{Figure~\ref{#1}}
\def \figurerefs#1{Figures~\ref{#1}}
\def \HeII{\hbox{He~{\sc ii}}}
\def \Rsun            {\hbox{$R_\odot$}}
\def \Rsolar          {\hbox{$R_\odot$}}
\def \Tsun            {\hbox{$T_\odot$}}
\def \Lsun            {\hbox{$L_\odot$}}
\def \Msun            {\hbox{$M_\odot$}}
\def \Rstar           {\hbox{$R_*$}}
\def \Rzero           {\hbox{$R_w$}}
\def \Rwind           {\hbox{$R_w$}}
\def \Rlam            {\hbox{$R_{\lambda}$}}
\def \Rw              {\hbox{$R_{w}$}}
\def \Blam            {\hbox{$B_{\lambda}$}}
\def \Slam            {\hbox{$S_{\lambda}$}}
\def \Teff            {\hbox{$T_{eff}$}}
\def \Rb              {\hbox{$R_{B}$}}
\def \Halpha          {\hbox{{\rm H}$\alpha$}}
\def \Hbeta           {\hbox{{\rm H}$\beta$}}
\def \Stromgren       {\hbox{Str\"{o}mgren}}
\def \RS              {\hbox{$R_S$}}
\def \Alam            {\hbox{$A_{\lambda}$}}
\def \vinf            {\hbox{$v_{\infty}$}}
\def \vesc            {\hbox{$v_{esc}$}}
\def \Mdot            {{\hbox{$\dot M$}}}
\def \MdotB           {{\hbox{$\dot M_B$}}}
\def \MdotR           {{\hbox{$\dot M_R$}}}
\def \Msunyr          {\hbox{$M_\odot\,$yr$^{-1}$}}
\def \kms             {\hbox{km$\,$s$^{-1}$}}
\def \VB              {{\hbox{$v_B$}}}
\def \Vir             {{\hbox{$V_{ir}$}}}
\def \VR              {{\hbox{$v_R$}}}
\def \vr      {{\hbox{$v_r$}}}
\def \Ne      {\hbox{$N_e$}}
\def \Np      {\hbox{$N_p$}}
\def \Ng      {\hbox{$N_g$}}
\def \percm3      {\hbox{cm$^{-3}$}}
\def \microns     {\hbox{$\mu$m}}
\def \kapzer    {\hbox{$\kappa_{\circ}$}}
\def \nuLnu       {\hbox{$\nu$ L$_{\nu}$}}
\def \lamLlam     {\hbox{$\lambda$ L$_{\lambda}$}}
\def \CaIV        {\hbox{[Ca\,{\sc iv}]}}
\def \third       {\hbox{\small{$\frac{1}{3}$}}}
\def \twothrd       {\hbox{\small{$\frac{2}{3}$}}}

\title{Constraints from ISO Data on the Velocity Law and Clumpiness
of WR~136\footnote{Based on observations with ISO, an ESA project with
instruments funded by ESA Member States (especially the PI countries:
France, Germany, the Netherlands, and the United Kingdom) and with the
participation of ISAS and NASA.}}

\author{R.~Ignace, M.~F.~Quigley, and J.~P.~Cassinelli }
\affil{Department of Astronomy, University of Wisconsin,
    Madison, WI 53706-1582}
\email{ignace@astro.wisc.edu, quigley@astro.wisc.edu, 
	cassinelli@astro.wisc.edu }

\begin{abstract}

Observations with the Infrared Space Observatory (ISO) SWS spectrometer
are used to constrain the velocity law and wind clumping of the
well-studied Wolf-Rayet (WR) star WR~136 (HD~192163) (WN6). Because
the free-free continuum opacity in WR~winds increases steadily with
wavelength in the IR, each point in the continuous spectrum may be
regarded as forming in a pseudo-photosphere of larger radius for longer
wavelength. Using this idea in combination with an analysis of the
Doppler-broadened widths of several \ion{He}{2} recombination lines,
we can derive information about the velocity law and clumpiness of the
stellar wind of WR~136. The observed line emission emerges from the region
exterior to the radius of optical depth unity in the free-free opacity,
corresponding to $v \gtrsim 0.3 \vinf$ for our shortest wavelength line.
The ISO observations provide the continuum shape, flux level, and seven
fairly strong \ion{He}{2} emission profiles. Adopting a $\beta$-law
distribution for the outflow velocity law, we compute the continuous
energy distribution and line profiles.  We find that there is a broad
range of $\beta$-values consistent with the continuum data if we also
allow the wind temperature to be a free parameter.  Interestingly,
the continuum data are found to constrain the wind to have fairly low
clumping values for the IR-forming region.  Using the continuum results
in conjunction with line profile modeling, the observational constraints
are best satisfied with a clumping factor of $D_c=<\rho^2>/<\rho>^2$ of
1--3 and a $\beta$-value of 2--3, although higher $\beta$ values are not
strongly ruled out for a modest wind temperature. The wavelength range of
our ISO data allows us to probe only the outer wind acceleration zone,
but in combination with radio observations, our finding that the wind
clumping is fairly small suggests that the clumping in the wind of WR~136
decreases with increasing radius.

\end{abstract}


\section{INTRODUCTION}

The Wolf-Rayet (WR) class of early-type stars has long been known for
having extreme stellar wind properties. In the visible band, broad
emission lines that form in the wind dominate the stellar spectrum
(Abbott \& Conti 1987).
The explanation of these extreme outflows has long posed severe problems
for stellar wind theories. In particular these stars are noted for their
``wind momentum problem'' (Barlow \etal\ 1981; Abbott \etal\ 1986;
Willis 1991), in which the momentum flux in the wind $\Mdot \vinf$
(where $\Mdot$ is the mass loss rate and $\vinf$ is the terminal speed
of the stellar wind) can exceed by more than an order of magnitude
the momentum flux $L_*/c$ (where $L_*$ is the stellar Bolometric
luminosity) of the radiation field purportedly responsible for the
outflows.  This situation is normally parametrized by the performance
number $\eta = \Mdot \vinf c / L_*$, which is of order 10 for WR~stars
(although there can be a wide range -- see Hamann \& Koesterke 1998).
In recent years the observational estimates of WR~wind momenta have
been reduced, partly from upward revisions of stellar luminosities (by
line blanketing, e.g., Hillier \& Miller 1999) and partly by accounting
for clumpiness in the winds (Hillier 1991; Hamann \& Koesterke 1998), yet
values of $\Mdot \vinf$ remain large ($\approx 10$, Nugis \& Lamers 2000).

Alternative wind-driving mechanisms have been proposed to explain the large
wind momenta, but several studies have indicated that the WR~winds can be
driven by radiative forces alone, without invoking other physics (such as
magnetic fields), through multi-line scattering of the stellar radiation
field (Lucy \& Abbott 1993; Springmann 1994; Gayley, Owocki, \& Cranmer
1995; Onifer \& Gayley 2003).  Multi-line scattering is possible because of
ionization gradients in the wind flow.  As photons escape from the inner
wind with an opacity characterized by a particular distribution of lines,
the photons encounter different sets of line opacities at larger
radii because of changes in the dominant ions.  Indeed a wide range
in ionization states as a function of radius appears to exist in the
WR~winds (Kuhi 1973; Schulte-Ladbeck, Eenens, \& Davis 1995; Herald \etal\
2000). Thus there can be a greater deposition of momentum by the stellar
radiation field to drive the wind, and an enhancement of
the wind momentum flux.

Although multi-line scattering can influence the mass-loss rates of some
O~stars, it tends to be much less important than for the WR~stars (Vink,
de Koter, \& Lamers 2000).  As we argue below, one of the consequences
of multi-line scattering in the wind is that the rate
at which the wind velocity increases with radius should be
shallower in WR~winds than in typical O~stars.
Observational constraints on the velocity laws of WR~stars should
thus provide a test of the multi-line scattering wind acceleration
process. Obtaining such information has been the motivation
for this observational study.

For early type stars, it is traditional to describe
the wind velocity distribution as a function of radius using the ``$\beta$
velocity law", which typically has the form,

\begin{equation}
v(r) =\vinf \left(1- \frac{\Rstar}{r}\right)^{\beta}.
	\label{eq:betalaw}
\end{equation}

\noindent This formula for the velocity distribution was derived with
$\bb = 0.5$ in one of the earliest of radiation-driven outflow models
by Milne (1926).  The same value was predicted in the line-driven
wind theory of Castor, Abbott, \& Klein (1975; CAK). In detailed
modeling of Of and OB supergiants by Groenewegen \& Lamers (1989),
values of $\bb =$0.7--0.9 were found. These higher values indicate a
shallower rise in the velocity distribution versus radius and are also
in better agreement with the predictions of the modified CAK (mCAK)
theory of Friend \& Abbott (1986) and Pauldrach, Puls, \& Kudritzki
(1986).  These mCAK models accounted for the finite size of the star
producing the wind.  They also led to predictions for \Mdot\ and \vinf\
that are in good agreement with observations of O~stars. The difference
in the \bb\ values in going from the CAK to the mCAK model arises from
the fact that the angular distribution of the driving radiation field
is different. For the very forward-peaked radiation field of CAK,
the acceleration is rapid, but for the somewhat more dispersed radiation
field associated with the finite disks of mCAK, the velocity law
is shallower. The multi-line scattering leads to an even
less strongly forward-peaked radiation field and to yet another step away
from a point-like radiation field. Thus it is to be expected that
the velocity law for WR~stars should have even higher values of \bb.

With the view that multi-line scattering can lead to a significant
enhancement of the winds, momentum transfer is no longer the major issue
in understanding the large mass-loss rates of WR~stars.  Instead there
is now a need for the opacity to adequately ``trap'' or prevent the
leakage of the photons so that multi-line scattering can indeed occur.
As the photons transmit some of their outward momentum to the wind,
they are red shifted. If there are gaps in the opacity versus wavelength
distribution, the photons can escape and thus contribute no further to
the wind acceleration. Since the winds of WR~stars require that photons
transmit about $\eta\times$ their momentum $h\nu_0/c$ to the wind,
a very large effective number of scatterings per photon is required.
For example, Gayley \etal\ (1995) show in their Figure 3, that a photon
will interact with approximately $\eta^2$ lines, and experience a net
redshift of $\Delta \lambda \approx\lambda_0\, (\eta\vinf/c)$.  A photon
that initially interacts with a EUV line near 600 \AA\ will tend to
drift about 660\AA\ redward.  Consequently, there can be few gaps in the
line opacity distribution, lest photons flood through opacity minima
to escape the wind before depositing sufficient momentum to the flow.
Hence the wind momentum problem has now come to be interpreted as an
``opacity problem'' (Gayley \etal\ 1995).

Rather little is known observationally about the radial distribution of
velocity in WR~winds. However, a good estimate for the wind terminal speed
\vinf\ can be made from forbidden line profiles (e.g., Barlow, Roche, \&
Aitken 1988; Willis \etal\ 1997; Morris \etal\ 2000; Ignace \etal\ 2001
[hereafter Paper~I]). This is because such lines are typically formed at
large distances of hundreds of stellar radii or more, where the wind has 
achieved its asymptotic expansion speed.  Because the forbidden lines
are optically thin and form in the constant velocity zone, they tend to
exhibit flat-topped profiles, the half-widths of which accurately determine
the wind terminal speed. 

Much less is known about the radial distribution of the velocity $v(r)$
deeper in the wind. There have been several discussions about the values
of \bb\ or the consequences of shallow velocity laws in WR stellar
winds. From theoretical modeling, Schmutz (1997) arrived at a \bb\
ranging from 3 to 8.  Monte Carlo radiation transport simulations by
Springmann (1994) and Gayley \etal\ (1995) also reveal extended wind
acceleration zones (not characterizable by a single $\beta$ value).
Support for these theoretical results comes from the study of binary
systems by (Auer \& Koenigsberger 1994). Also, L\'{e}pine \& Moffat
(1999) have derived large \bb\ values from observations of the regular
motion of small emission peaks or ``migrating features'' that appear
in WR He II emission lines. An interesting structural effect of large
\bb\ values was discussed by Ignace, Cassinelli, \& Bjorkman (1996). They
found that velocity laws with modestly large values of $\beta$ (such
as $\bb=3$) are more likely to form wind-compressed zones as a result
of stellar rotation.  This effect was used to explain why some WR~stars
show intrinsic polarization (e.g., Harries, Hillier, \& Howarth 1998).

To study the velocity law in WR~stars further, we obtained Infrared Space
Observatory (ISO) satellite observations, and in this paper we describe
results from the best data set, which corresponds to WR~136. We use
observations of both the IR free-free continuum and emission profiles
of \ion{He}{2} recombination lines.  In \S 2, we discuss in general the
continuum and line diagnostics of the density and velocity distributions
in WR stars. In \S 3, the description of the data obtained with the ISO
satellite and its analysis to infer \bb\ is presented.  The results of
the study are discussed in \S 4.

\section{INFRARED DIAGNOSTICS FOR DENSE STELLAR WINDS}

Relative to the flows from other early-type stars, the WR~winds are
denser and they produce spectra that are dominated by emission lines. The
understanding of these spectra involves three key elements: (a) In
the optical, a major continuum opacity is electron scattering, which
is gray but involves complicated thermalization effects (Mihalas 1978)
(b) In the UV and EUV spectral ranges, strong line blanketing both blocks
the flow of radiation and drives the wind.  (c) In the IR through to the
radio spectral region, it is the increasingly large  free-free opacity
that accounts for the IR and radio continuum excesses.  The advantage of
using this long wavelength spectral region to study the wind structure
is that the continuum opacity increases monotonically with wavelength.
For wind temperatures of a few times $10^4$~K that are applicable
to radii where the free-free continuum forms, the opacity varies
as $\lambda^3$ in the near IR, gradually shifting to a $\lambda^2$
dependence at longer wavelengths.  In the latter case, the free-free
opacity is so great that the emergent radiation originates at large
radii where the flow has reached terminal constant expansion-velocity.
The free-free opacity is lower in the IR, and this permits radiation to
emerge from the parts of the wind where the flow is still accelerating.
In particular, ISO observations provides us with the spectral data
that allows us to diagnose an interesting segment of the wind velocity
law. There are two complementary approaches that we employ: (a) a study
based on the continuous energy distribution, and (b) a study of emission
line profiles formed by recombination.  The latter method is, in fact,
related to the first in that the continuos opacity of the wind affects
the range of depths over which we receive line emission, and thus can
have a strong influence on profile strengths and shapes.  Consequently,
we consider the continuum first, and then proceed to the emission
line profiles.  Many of the issues relevant to the interpretation of IR
observations of WR~stars have been discussed in investigations similar
to ours, such as Hillier, Jones, \& Hyland (1983) and Hillier (1987ab)
in detailed studies of WR~6 ($=$ HD~50896 $=$ EZ~CMa).

\subsection{IR Continuum Slope Methods}

The classic approach to studying WR density and velocity distributions is
to consider the continuum slope.  Wright \& Barlow (1975) and Panagia \&
Felli (1975) were the first to demonstrate that the spectral slope should
vary as $f_\nu \propto \nu^{0.6}$ (or $f_\lambda \propto \lambda^{-2.6}$)
for a spherical wind in constant expansion-velocity and thick in the
free-free continuum.  There are several informative scalings that came
from these early studies and subsequent considerations.  For example,
Wright \& Barlow showed that in the constant expansion-velocity case,
the continuum emission is equivalent to a volume integral over the
emissivity $j_\nu \propto \rho^2$ with a lower bound $R_{\nu,{\rm vol}}$
determined by the radius at which the line-of-sight free-free optical
depth is $\tau_{\nu,{\rm ff}} = 0.244$.  This optical depth is somewhat
smaller than the $2/3$ value associated with the Eddington-Barbier
result for plane-parallel stellar atmospheres.  Adopting an ``equivalent
photosphere'' approach, Hillier \etal\ (1983) showed that for the
observed monochromatic luminosity with $L_\nu = 4\pi\,R^2_{\nu,{\rm
surf}}\,\pi\,B_\nu(T_{\rm e})$, where $B_\nu$ is the Planck function
at the wind temperature, the monochromatic radius $R_{\nu,{\rm surf}}$
corresponds to a radial depth of $\tau_{\nu,{\rm ff}} = 0.05$. This
small optical depth means that to understand the monochromatic
luminosity, it is essential to account for the spatially extended
nature of the wind continuum formation region.

As in Wright \& Barlow (1975) and Hillier \etal\ (1983), let
us compare the line-of-sight optical depths for different wind velocity
laws.  For the geometry depicted in Figure~\ref{fig1}, the free-free optical
depth along a ray that passes through the envelope at impact parameter $p$
(normalized to $R_*$) is given by

\begin{equation}
\tau_{\nu, {\rm ff}} (p,z) = R_*\,K_{\rm ff}(T,\nu)\,\int_z^\infty\, n_{\rm i}
	(x)n_{\rm e}(x) \,dz',
	\label{eq:tff}
\end{equation}

\noindent for $x= \sqrt{z^2+p^2} =r/R_*$ the normalized stellar radius,
$z=Z/R_*$ the line-of-sight coordinate toward the observer, $n_{\rm i}$
the ion number density, $n_{\rm e}$ the electron number density, and
$K_{\rm ff}(T,\nu)$ a factor that isolates the temperature and frequency
dependence of the free-free opacity.  This opacity factor is given in
Wright \& Barlow as

\begin{eqnarray}
K_{\rm ff}(T,\nu) & = & 3.7\times 10^8\,Z_{\rm i}^2\,g_\nu(T)\,T^{-1/2}\,\nu^{-3}\nonumber\\ 
& & \times \left(1-e^{-h\nu/kT}\right)\;\;{\rm cm^{5}\;K^{1/2}\;Hz^3},
\end{eqnarray}

\noindent where $Z_{\rm i}$ is the rms ion charge, $g_\nu(T)$ is the Gaunt
factor, $T$ is in K, and $\nu$ is in Hz.  We assume that $T$
is constant throughout the continuum-formation portion of the wind, and
its value is treated as a free parameter, for which we
know some limiting values.

We assume a spherical radial wind and use the dimensionless radius $x=r/R$
and velocity $w=v/\vinf$.  From the continuity equation, the radial
distribution of density is given by

\begin{equation}
\rho = \frac{\Mdot}{4\pi\,r^2\,v(r)} = \frac{\rho_0}{x^2\,w(x)},
\end{equation}

\noindent where the scale factor is

\begin{equation}
\rho_0= \frac{\Mdot}{4\pi\,R_*^2\,\vinf},
\end{equation}

\noindent and the normalized wind velocity distribution is

\begin{equation}
w(x) = \left(1-\frac{b}{x}\right)^\beta,
	\label{eq:normvlaw}
\end{equation}

\noindent for 

\begin{equation}
b = \left[ 1 - \left(\frac{v_0}{\vinf}\right)^{1/\beta} \right]
\end{equation}

\noindent Equation~(\ref{eq:normvlaw}) differs from
equation~(\ref{eq:betalaw}) in that we allow for the parameter $b \le
1$ and provides for a finite initial value $w_0=v_0/\vinf$ at the base
of the wind where $x=1$.  To evaluate the free-free optical depth in
equation(\ref{eq:tff}), the electron and ion densities are related to
the mass density as $n_{\rm e} = \rho / \mu_{\rm e} m_H$ and $n_{\rm i}
= \rho / \mu_{\rm i} m_H$, for $\mu_{\rm e}$ and $\mu_{\rm i}$ the mean
molecular weight per free electron and free ion, respectively.

For illustrative purposes, let us compare two simple cases: constant
expansion-velocity (i.e., $\beta=0$) and a $\beta=1$ velocity law
involving typical wind acceleration versus radius.  The comparison
is motivated by the fact that we seek to use the observed spectral
energy distribution to infer information about the wind velocity law.
With these two values of \bb\ we gain insight into how the velocity law
influences the observables, and  conversely what our ISO observables
can tell us about the velocity distribution.

For the constant expansion-velocity case, the line-of-sight optical depth 
$\tau_{\rm ff}(x)=\tau_{\rm ff}(z,p=0)$ becomes

\begin{equation}
\tau_{\nu, {\rm ff}}(x) = \frac{1}{3}\,\tau_{\rm c} (T,\nu)\, x^{-3},
\end{equation}

\noindent so that a specified optical depth is obtained at a radius
$x_{\tau} = (\tau_{\rm c}/3\tau)^{1/3}$, where the constant $\tau_{\rm
c}$ collects all of the various wind, atomic, and frequency dependencies
in a single parameter.  The total radial optical depth down to the wind
base at $x=1$, is $\tau_{\rm c}/3$.  The emergent intensity from any
point along the line-of-sight is $B_\nu\,\exp(-\tau_{\nu, {\rm ff}})$.

For the $\beta=1$ velocity law, the integral for the optical depth is again
analytic, giving

\begin{equation}
\tau_{\nu, {\rm ff}}(u) = \frac{\tau_{\rm c}(\nu)}{b^3}\,\left[ \frac{1}{1-bu}
	- (1-bu) + 2\ln(1-bu) \right].
\end{equation}

\noindent where $u=1/x$.  The total line-of-sight optical depth to the
wind base is $\tau_{\rm c}\, b^{-3}[b(2-b)/(1-b)+2\ln(1-b)]$.

In Figure~\ref{fig2}, both the optical depth and the emergent intensity
$I_\nu$ are plotted against radius for these two cases.  In this example
both $K_{\rm ff}$ and $\rho_0$ have been held constant.  There are four
important points to note.  (a) The total line-of-sight optical depth is
greater for larger $\beta$, because at every radius, the flow will be
slower and hence denser.  (b) A fixed optical depth occurs at larger
radius for larger $\beta$. (c) Thus the observed free-free emission
emerges from larger radius for larger $\beta$, leading to a greater
monochromatic free-free luminosity $L_\nu$.  And (d) for a constant
source function $B_{\nu}(T)$, intensities emerge from contributions over a
broad range of radii.  When evaluating the total flux from the unresolved
source, the net effect of larger values in $\beta$ is to steepen the wind
density distribution, which in turn leads to a steeper continuum spectrum.

Instead of the ``forward problem'' of working from an assumed wind velocity
law toward a continuum slope, the relation between the wind
expansion and observed continuum emission can be addressed with an
``inverse approach''.  Cassinelli \& Hartmann (1977) considered the
relation between the observed spectral slope and the density and
temperature variations assuming power law forms, with $\rho \propto
r^{-q}$ and $T \propto r^{-m}$.  Using the effective radius concept,
they derived a relation between the power-law slope $\nu f_\nu =
\lambda f_\lambda \propto \lambda^{s}$ and the $q$ and $m$ indices
as given by

\begin{equation}
s = \frac{6q- \frac{5}{2}m-7}{-2q+\frac{3}{2}m+1}.
	\label{eq:s}
\end{equation}

\noindent Taking the wind to be isothermal with $m=0$,
equation~(\ref{eq:s}) reduces to $s = (6q-7)/(1-2q)$.  For a spherical
wind, inferring $q$ for the density law is equivalent to determining
the {\it shape} of the wind velocity distribution, although not the 
values of the velocity. Furthermore, given the
range in wavelengths of the IR observations, the shape of the derived velocity
distribution is determined over the corresponding range in continuum formation
radii. More generally, the spectral slope will depend not only on the
density distribution, but also on the temperature gradient,
electron scattering, bound-free opacity,  and wind ionization variations.
The main point of this discussion is that the wind velocity
law can control the continuum spectral slope and may thus be inferable
from observations.

\subsection{The IR Emission Line Profile Method}

\begin{figure}
\plotone{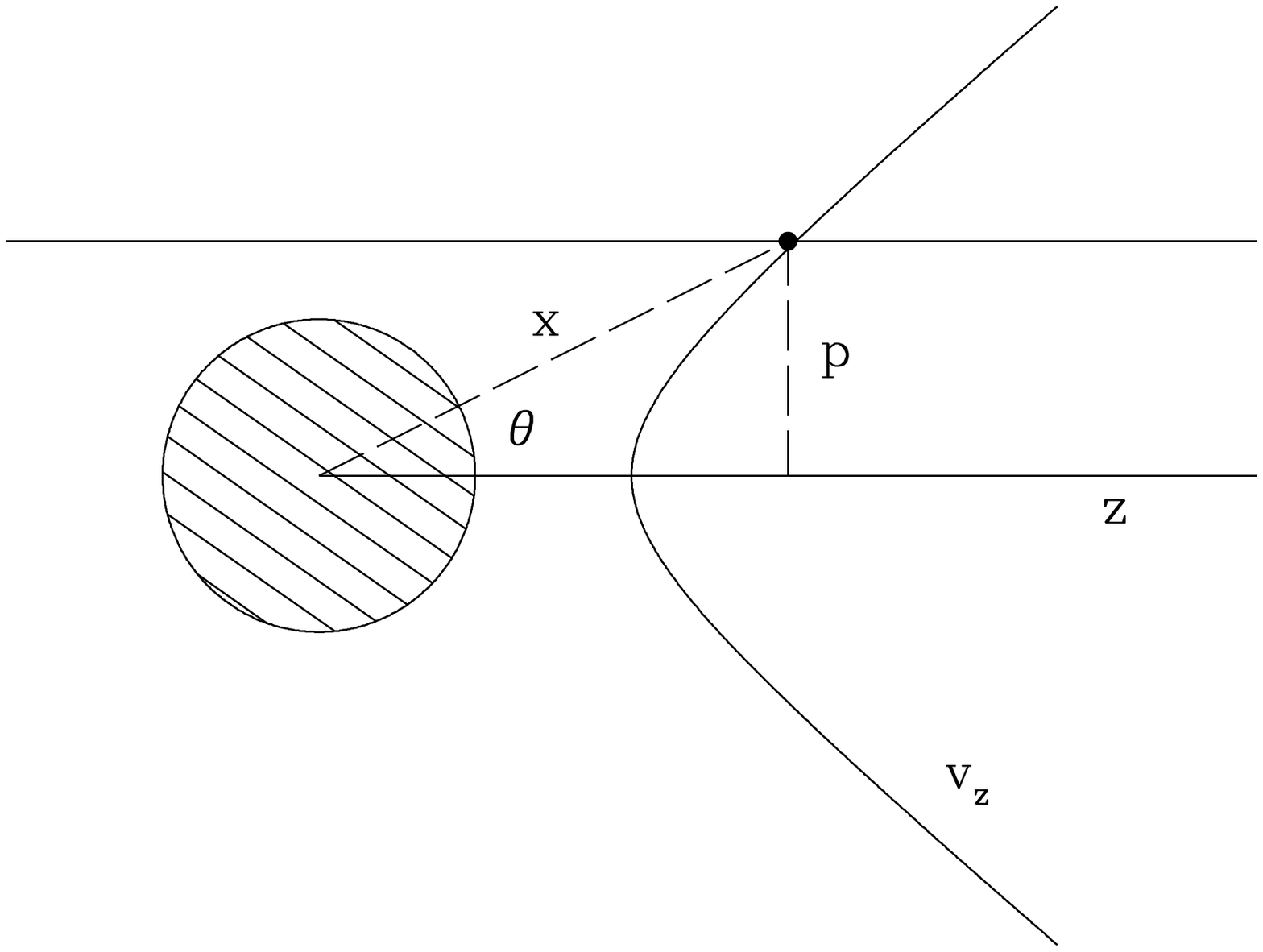}
\caption{The geometry used in our calculations.  All lengths are
normalized to the stellar radius $R_*$.  The
observer is to the right.  The observer coordinates
are the cylindrical system $(p,\alpha,z)$, with azimuth $\alpha$ not shown.
The star has spherical coordinates $(r, \theta, \phi)$.  With the
assumption of spherical symmetry, we chose $\theta$ measured from
the $z$-axis as indicated.  Also shown is a
contour for an isovelocity zone with $v_{\rm z}=$ constant and a
ray passing through the envelope from the observer direction.  At
the frequency in the line corresponding to the Doppler shift $v_z$,
the contribution to the observed emission along this ray includes
line emission from the indicated point and continuum emission from
both fore and aft of this point, along with the associated attenuations. }
\label{fig1}

\end{figure}

\begin{figure}
\plotone{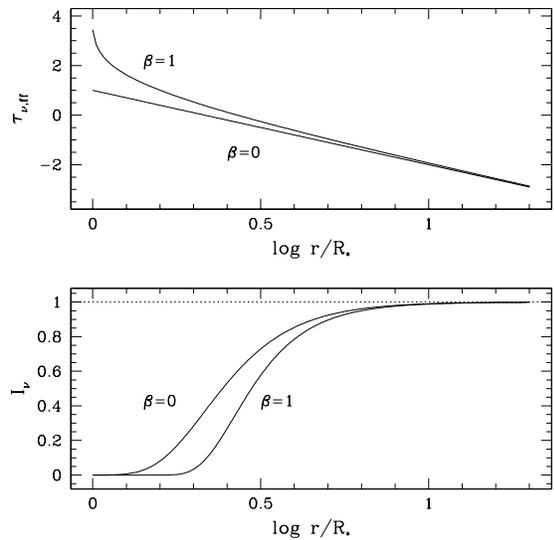}
\caption{Illustrates the influence of the velocity law on the
optical depth (top) and emergent intensity (bottom) with $I_\nu =
B_\nu\,\exp(-\tau_{\nu,{\rm ff}})$ along the line-of-sight to the star.
The two cases shown are for $\beta=0$ (constant expansion-velocity) and
$\beta=1$.  Unlike a static exponential atmosphere, the extended envelope
effects of an expanding wind lead to an attenuation profile that is not
especially sharp as a function of radius.
} \label{fig2}

\end{figure}

Our data set
contains a number of IR recombination lines of \ion{He}{2}.  
The strong lines observed in our spectra are for transitions
involving levels in the range 6 to 16.
Assuming pure recombination lines and accounting for free-free opacity,
we model the line profiles using standard Sobolev theory in spherical
symmetry (e.g., Mihalas 1978).  The total emergent intensity arriving at the
Earth along a ray at $p$ (see Fig.~\ref{fig1}) is

\begin{equation}
I_\nu^{\rm emer} (p) = I^{\rm emer}_{\nu,{\rm c}} + I^{\rm emer}_{\nu,l}
\end{equation}

\noindent where the first term on the RHS is the emergent intensity of
the underlying continuum emission and the second term is the emergent
intensity of line emission.  Both the line and continuum emission are
computed at each frequency in the line profile.  In Sobolev theory,
the line emission at a given frequency arises from an ``isovelocity
zone'', which is a region of constant Doppler shift in the flow as
perceived by the observer.  Figure~\ref{fig1} shows an example of an
isovelocity surface.  The continuum emission, however, is determined by
a volume integral over the whole envelope.  The continuous opacity can
attenuate the line emission; moreover, the line opacity can diminish
the net continuum emission.

It is convenient to define four optical depths.  The first three are for
the continuos opacity: $\tau_{\rm fore}$, $\tau_{\rm aft}$, and $\tau_{\rm
tot}$.  Of these the first is the optical depth that attenuates the line
emission (i.e., by matter that lies on the near side of the isovelocity zone 
with respect to the observer).  The second does not (i.e., matter  
on the rear side of the
isovelocity zone).  The third is simply the sum, $\tau_{\rm tot}=\tau_{\rm
fore}+\tau_{\rm aft}$.  The fourth optical depth is the Sobolev optical
depth $\tau_S$ associated with the line opacity.  This is given by

\begin{equation}
\tau_S = \frac{\kappa_l \rho \lambda}{| dv_{\rm z}/dz |} = 
	\tau_l\,\left[ x^4\,w^2(x)\,\frac{dw_{\rm z}}{dz} \right]^{-1}
\end{equation}

\noindent where the line-of-sight velocity gradient is

\begin{equation}
\frac{dw_{\rm z}}{dz} = \mu^2\,\frac{dw}{dx}+(1-\mu^2)\,\frac{w}{x},
\end{equation}

\noindent where $\mu=\cos \theta$, and the scale constant $\tau_l$ is

\begin{equation}
\tau_l = \frac{\kappa_l\,\rho_0\,\lambda\,R_*}{v_\infty}.
\end{equation}

Using these optical depth definitions, the emergent 
continuum and line intensities are given respectively by

\begin{equation}
I_{\nu,{\rm c}} = \int^{\tau_{\rm tot}}_{\tau_{\rm fore}} \, B_\nu(T)\,
	e^{-\tau}\,e^{-\tau_S}\,d\tau + \int^{\tau_{\rm fore}}_0
        \, B_\nu(T)\,e^{-\tau}\,d\tau
\end{equation}

\noindent and

\begin{equation}
I_{\nu,l} = S_{\nu,l}\,(1-e^{-\tau_S})\,e^{-\tau_{\rm fore}}.
	\label{eq:lineinten}
\end{equation}

\noindent If we assume that the wind is isothermal, or at least
approximately constant over the radii giving rise to the bulk of the
line and continuum emission, the intensity for the continuum reduces to

\begin{equation}
I_{\nu,{\rm c}} = B_\nu(T_{\rm w})\,
        \left(e^{-\tau_{\rm fore}}-e^{-\tau_{\rm tot}}\right)\,e^{-\tau_S} + 
        B_\nu(T_{\rm w})\,\left(1-e^{-\tau_{\rm fore}}\right),
	\label{eq:continten}
\end{equation}

\noindent for $T_{\rm w}$ the wind temperature.  If the further assumption
is made that the departure coefficients for the upper level populations
are nearly unity, one has that $S_{\nu,l}=B_\nu(T_{\rm w})$, and the
total emergent intensity for line and continuum emission, including the
relevant absorptions, along a ray at $p$ compactly reduces to

\begin{equation}
I_{\nu}(p) = B_\nu(T_{\rm w})\,\left[1-e^{-(\tau_{\rm tot}+\tau_S)}\right]
	\label{eq:simplified}
\end{equation}

\noindent Clearly, outside the line frequencies, the calculation for the
continuum emission is given by the same expression only with $\tau_S=0$. We
ignore any bound-free contribution to the opacity (see the discussion of \S
4), but note that the bound-free emissivity at IR wavelengths also scales as
$\rho^2$ so that the form of equation~(\ref{eq:simplified}) would continue
to hold.  Electron scattering is also ignored, but it becomes more important
at shorter wavelengths where the free-free opacity is smaller.
Equation~(\ref{eq:simplified}) implies that in the line, the intensity is
always less than that in the neighboring continuum. In such a case the only
way that a line will appear in emission is if the integration of the
intensity over the apparent disk of the envelope is larger at line
wavelengths. Thus an emission line profile provides us with information about
the velocity distribution somewhat beyond the radii at which the neighboring
continuum is formed.

\begin{figure}
\plotone{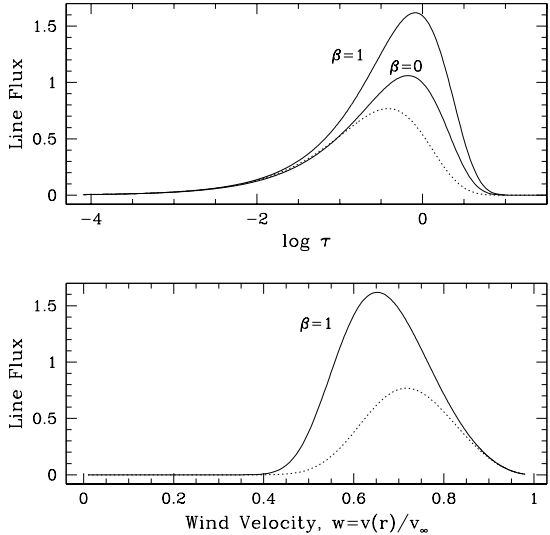}
\caption{Illustrates the influence of the velocity law on the emergent
line contribution function along the line-of-sight to the star plotted
against the continuum optical depth (top) and wind velocity in the wind
(bottom).  At top, the two solid curves are optically thin lines for the
indicated $\beta$-laws.  The dotted line is for an optically thick line
with $\beta=1$.  The effect of large line optical depth is to reduce the
amount of emergent line flux and shift the peak location to larger radii.
At bottom, just the $\beta=1$ curves are plotted against the normalized
wind velocity.  The range of velocities over which line flux contributions
arise is seen to be fairly broad.  In this case, for the optically thin
line, the peak in the line flux contribution occurs around $v=0.65\vinf$
and has a full width of $\approx 0.24 \vinf$.  } \label{fig3}

\end{figure}

We are interested in using the line data to learn about the wind
velocity distribution.  Figure~\ref{fig3} shows the emergent flux of
line emission, given by $j_\nu \, \exp (-\tau_{\nu,{\rm ff}})\,dV$,
for $j_\nu$ the line emissivity and $dV$ a differential volume element,
along the line-of-sight toward the center of the star.  The upper panel
is a plot against continuum optical depth, and the lower is against the
normalized wind velocity.  The two solid curves at top are for explicitly
optically thin lines, contrasting the constant expansion-velocity case
with $\beta=0$ against the $\beta=1$ case. Also plotted is a dotted line
for an optically thick line with $\beta=1$.  From Figure~\ref{fig3},
we may conclude that the bulk of the line emission emerges exterior to
the radius where $\tau_{\nu,{\rm ff}}=1$.  In the bottom panel, only
the thin (solid) and thick (dotted) line cases for $\beta=1$ are shown,
since there is no variation of $v$ with $r$ for $\beta=0$.  We see that
the line emission originates from a fairly broad range of physical wind
velocities.  At longer wavelengths, where the free-free opacity is larger,
the peak of the flux contribution will move to higher wind velocities
resulting in emission lines of increasing breadth.

To obtain the total line flux at a given line frequency (or velocity
shift $v=v_{\rm z}$) requires an integration over impact parameter for
the corresponding isovelocity zone, as given by

\begin{equation}
f_\nu = \frac{2\pi\,R_*^2}{d^2}\,\int_{w_{\rm z}}\,I_\nu (p)\,p\,dp,
	\label{eq:fline}
\end{equation}

\noindent where $d$ is the distance to the star.  For thin lines
the integrand of equation~(\ref{eq:fline}) scales with $\rho^2 r^2
\exp(-\tau_{\nu,{\rm ff}}) \propto x^{-2} w^{-2} \exp(-\tau_{\nu,{\rm
ff}})$.  The width of the line profile depends on the wind expansion in
the region where the line forms.  Since there is a finite range of radii
that primarily contribute to the line emission, the overall width of the
line is set by the velocity at the radius where $\tau_{\rm ff}\approx 1$.
Line emission from smaller radii is strongly absorbed, and that from
larger radii is small owing to the diminishing emission measure.

In this discussion of the line radiative transfer, the forward approach
has been emphasized.  As was the case for the continuum, it is possible
to use an inverse approach to analyze emission lines, if certain
simplifying assumptions hold. The inverse approach can allow one to
directly determine the wind velocity law (Cannon 1974; Brown \etal\ 1997;
Ignace \etal\ 1998ab).  The primary assumptions are that the lines be
optically thin, the wind be spherically symmetric, and that there be a
relatively distinct photospheric radius.  Unfortunately, consideration
of the S/N, difficulties with line blends, non-thermal broadening (e.g.,
from wind instabilities), and the radial extension of the free-free
continuum formation zone, have led us to conclude that the application of
these inverse techniques to the observed \ion{He}{2} profile shapes in
WR~136 will not be useful.  The method of Ignace \etal\ (1998b) appears
to show some promise.  It uses the distribution of total line intensities
for several thin lines, and the total emission for each line depends on
the emission measure.  Thus some of the effects that severely bias the
inversion of line profile shapes are averaged out.  However, there are
not enough suitable lines in our dataset to attempt this inversion.

Returning to the forward approach, a key point in modeling
the IR spectrum of dense winds is that both the line and continuum optical
depths scale with the square of the density.  The assumption that the
source function for the line and continuum are the same ensures that
there is no net absorption feature in the profile.  So the situation
results in rather symmetric emission profiles, which initially is
somewhat counterintuitive, since the absorption of line emission by
the continuos opacity varies strongly from the near side of the star
(corresponding to blueshifted line frequencies) to the far side (for
redshifted line frequencies).

A counter example is provided by wind X-ray emission lines
that are expected to be asymmetric in the standard paradigm of
distributed wind shocks (Ignace 2001; Owocki \& Cohen 2001; Ignace
\& Gayley 2002).  For X-ray lines, the emissivity scales with
$\rho^2$, and the line emission suffers photo-absorption by the
dominant ``cool'' wind component.  Asymmetry in the profile shape
results because the continuous opacity makes no contribution to
the emission, and such asymmetric morphologies are observed in the
{\it Chandra} spectra of $\zeta$~Pup (Cassinelli \etal\ 2001).
Indeed in the case at hand, we have examined the line emission
contribution alone (i.e., with attenuation by free-free absorption)
and find it to be strongly asymmetric with blueshifted peak emission,
just like the theoretical X-ray lines.  However, the absorption of the
continuum is also asymmetric, with greatest absorption occurring at
blueshifted frequencies.  This is why the net profile recovers a fairly
symmetric appearance.

To show profile effects explicitly, an analytic expression for the profile
shape can be derived in the case of a constant expansion-velocity flow,
if the free-free optical depth is quite large.  In fact, the profile
shape in this limit has previously been derived by Hillier \etal\ (1983),
but the result appears there as an unnumbered equation (between their
eqs.~[13] and [14]), and examples of the profile shapes are not shown.
Those authors were more concerned with line equivalent widths.  (In their
appendix a derivation for the equivalent width is presented for a wind in
homologous expansion, with $v\propto r$.) The solution for this special
case is particularly apt to the problem at hand, so we provide details
for the line profile derivation in the appendix.  Results are shown in
Figure~\ref{fig4}, where a sequence of line profiles are plotted as the
ratio of line to continuum optical depth scales, $\tau_l/\tau_{\rm c}$,
is varied.  The profile is rather ``bubble-shaped'' and is identically
symmetric.  Of course, relaxing any of our assumptions will, in general,
allow the profiles to develop asymmetries; however, the bubble-shape
profiles illustrated by the constant expansion-velocity case are similiar
to those actually observed in WR~136, as we describe next.

\section{{\it ISO} OBSERVATIONS OF WR~136}

We have obtained {\it ISO} observations of the prototype
WN6 Wolf-Rayet star WR~136 (HD~192163).  As a member of the Cyg~OB1
association, the distance to this star is well-determined ($d= 1,820$~pc;
Lundstr\"{o}m \& Stenholm 1984). WR~136 is unusual among 
WN6~stars in that it has a substantial abundance of hydrogen,
$\sim 12\%$ by mass (Crowther \& Smith 1996; Hamann \etal\ 1994). Line
and continuum spectra of WR~136 were obtained with the SWS spectrometers
aboard the {\it ISO} satellite as described in Table~\ref{tab1}.
Automated preliminary processing of the data was done using version 7
of the standard pipeline package (OLP-7), and subsequent processing
was done as described in detail in Paper~I. In this section we present
the continuum slopes and line profile widths that can be used to
derive density and velocity information for the wind of the star.

\subsection{The Continuum Flux Distribution}

Continuum measurements were made over the range of 2.8 to 10~microns,
beyond which the S/N ratio was insufficient to reliably determine a
continuum.  The spectrum obtained with the SWS01 spectrometer is displayed
in Figure~\ref{fig5}, and this has been dereddened using the algorithm
of Whittet (1992), with values for $A_V$ and $E_{B-V}$ given by van der
Hucht (2001).  A power-law fit was made to the continuous flux distribution,
as shown in Figure~\ref{fig6}, that yields a power-law index of

\begin{equation}
\gamaF= \frac{d \log f_\lambda}{d \log \lam}= -2.87.
\label{Eq:gammaF}
\end{equation}

\noindent This value is to be compared with a slope of $-2.6$ expected
from a constant expansion wind at long wavelengths (Wright \& Barlow
1975).  The steeper observed value suggests that either the wind is
accelerating, the temperature is varying, or both.  However, we find
that the power law is somewhat weakly constrained, especially by the low
quality data at longer wavelengths.  The $1\sigma$-confidence interval for
acceptable power-law indices is quite asymmetric, with a range of $-2.75$
to $-2.9$ that applies to the better quality short wavelength spectrum
(the $-2.87$ is the minimum $\chi^2$ value).  This range is consistent
with spectral indices reported by Nugis, Crowther, \& Willis (1998), who
quote a value of $-2.75\pm 0.02$ in the range of 12~\microns\ to 6~cm,
and a value of $-3.15\pm 0.35$ in the range of 12--25~\microns.

Using the range of power laws from our ISO data, we can estimate the density
variation over the range of radii in which the continuum forms for these
wavelengths.  In the power-law approach of Cassinelli \& Hartmann (1977)
described in \S 2.1, the $s$ in equation~(10) is related to \gamaF\ via $s =
\gamaF+1$, which in our case ranges from $-1.75$ to $-1.9$. Assuming an
isothermal wind with $m=0$, the density power law is found to be
$q=(s+7)/(2s+6)$, which ranges from 2.1 to 2.3.  But of course, if $\rho
\propto r^{-q}$, and $\rho \propto r^{-2} v^{-1}$, then
$v \propto r^{q-2}$, which in this case results in $v\approx
r^{0.2}$. This is a fairly shallow slope, and consequently the continuum
data, based on this crude analysis, is likely arising from rather far out in
the wind acceleration zone, where the wind is starting to approach constant
expansion-velocity.  

In terms of a standard wind $\beta$~law like equation~(\ref{eq:normvlaw}),
one can make a further identification.  Assuming large radius with $r\gg
R_*$, the velocity gradient is $d\ln v/d\ln r \approx \beta b R_*/r$.
With $v \propto r^{0.2}$, we have that $\bb \approx 0.2 r/bR_*$,
which implies that $v \approx \vinf (1- \bb bR_*/r) \approx 0.8\vinf$
regardless of the $\beta$-exponent. This exercise suggests that our data
probe mainly the outer wind acceleration.

\subsection{The \ion{He}{2} Line Profiles}

The high resolution SWS06 observing mode was used to obtain recombination
line data with resolving powers ranging from $\lam/\Delta\lam
\approx 1300$ to 2500 over wavelengths of 2.8 to 5.0~microns,
respectively. We use the seven best \ion{He}{2} recombination lines in
our study. Figure~\ref{fig7} shows the six lines used for determining
the line half-widths (the seventh, 14--11 is too noisy to obtain a
reliable width, but is adequate for determining an equivalent width). Line
identifications with comments are given in Table~\ref{tab2}.  Atomic data
and measurements for these lines are provided in Table~\ref{tab3}. The
HWHMs for the lines are around 1000~\kms, and the instrumental broadening
is about 100~\kms.  Since the emission profiles are somewhat Gaussian,
estimating the effect of instrumental ``smearing'' is relatively
straightforward.  The convolution of two Gaussians yields another
Gaussian.  The width of the resultant Gaussian is given by the rms of
the widths for the two input Gaussians, in this case the emission line
and the instrumental smearing.  Thus, instrumental broadening affects
our HWHM determinations at about the 1\% level, and so we have ignored it.

Using the line width data, we can make another estimate of the wind
velocity law in WR~136.  HWHM values were measured for the {\em red}
wing of the six \ion{He}{2} recombination lines and are plotted against
wavelength in Figure~\ref{fig8}.  (HWHMs measured from the blue wing
are systematically less by about 100~\kms, which we attribute to the
influence of line absorption.)  A power-law fit was made to this data,
yielding a power-law index of

\begin{equation}
\gamav= \frac{d \log {\rm HWHM}}{d \log \lambda} = 0.22 \pm 0.07,
\label{Eq:gammaw}
\end{equation}

\noindent To infer information about $v(r)$, a power-law index slope in
radius is needed.  For this, one could assume that the line emission
originates from the vicinity of the pseudo-photosphere of the continuum
(i.e., around $\tau_{\nu,{\rm ff}} =1$), since the line emission interior
to this region will be strongly absorbed.  But without knowing the wind
velocity law {\it a priori}, it is impossible to compute the radius of
optical depth unity in the free-free opacity.  We can at least estimate
the wavelength dependence of the radius as follows.

From the observed continuum emission, we can define the surface
equivalent radius that would provide the observed flux, with

\begin{equation}
R_{\lambda,{\rm surf}} = d\,\sqrt{\frac{f_\lambda}{\pi B_\lambda}}.
	\label{eq:Rlamsurf}
\end{equation}

\noindent Already from \S 2.1, we have argued that this radius
$R_{\lambda,{\rm surf}}$ will be larger than the radius where
$\tau_{\nu,{\rm ff}}=1$.  Even so, the overall wavelength dependence
of $R_{\lambda,{\rm surf}}$ should roughly mimic that of the radius
of optical depth unity, namely that at longer wavelengths, the two
radii will increase in size in a manner that is roughly proportional.

With $f_\lambda$ and $B_\lambda$ taken as known, the run of
$R_{\lambda,{\rm surf}}$ with wavelength can be determined, and we
find that $\gamma_R=d\log R_\lambda/d\log \lambda = 2+0.5\gamaF$ or
about 0.6.  So we may infer that $d\log v/d\log R_{\lambda,{\rm surf}}
= \gamav/\gamma_R \approx 0.4$, which is consistent with the estimate
(of 0.2) derived from \S 3.1 based purely on the continuum slope data.
Of course, these two approaches (continuum slope and line width)
are not entirely independent, since $\gamma_R$ is related to \gamaF.
Nevertheless, the line-width data is independent, and so the consistent
results support the argument that we are observing the outer wind
acceleration at these wavelengths.

\subsection{Combined Analysis of Continuum and Lines}

We have written a code to compute both free-free emission and
recombination line profiles using the formalism of \S 2.2.  We use this
code to reproduce the continuum fluxes and line profile shapes for our
best \ion{He}{2} lines.  The goal is to simultaneously constrain the
velocity law $\beta$, the characteristic wind temperature $T_{\rm w}$,
and the wind clumping.

\subsubsection{Clumping Factor Effects}

Clumpiness, now a well-recognized attribute of many WR~stars (e.g.,
Moffat \etal\ 1988; Hillier 1991; Nugis, Crowther, \& Willis 1998;
Hamann \& Koesterke 1998), is included in our modeling.  We define the
clumping factor parameter as

\begin{equation}
<\rho^2>\; = \Dclump\ <\rho>^2 .
	\label{Eq:Ddef}
\end{equation}

\noindent The wind clumping is assumed to occur on small scales, so as
to produce the same emission measure as a smooth wind but with a
smaller \Mdot.  The clumping may occur on a range of scales, but for
our theoretical approach, these scales must be small enough that the
Sobolev approximation is not invalidated.  In essence, this means that
the radiative transfer may still be described by integral expressions,
in contrast to discrete sums (Hamann \& Koesterke 1998b).

\subsubsection{Assumed Wind Parameters}

We consider the terminal speed to be known from Paper~I, at
$\vinf=1490~\kms$.  For the mass-loss rate, Table~\ref{tab4} lists
several values for WR~136 as taken from the literature; the method of
evaluation; and whether the estimates assumed smooth or clumpy winds.
The first four values assumed an underlying smooth wind.  For the last
value, Nugis \etal\ (1998) used radio observations to derive \Mdot\
values using a clumped wind model, and found that at the
large radii where the radio emission is produced, the wind of WR~136 is
consistent with no clumping (i.e., $\Dclump \approx 1$).  On the
other hand, Crowther \& Smith (1996) conclude in favor of clumping in the
wind of WR~136 because otherwise their models overestimate the electron
scattering wings of strong optical/NIR lines.  That these two different
methods yield different clumping predictions must reflect on the
fact that the methods sample different radial locations in the wind.
It may not be surprising that clumping can be a function of radius
(e.g., as in the theoretical models of Runacres \& Owocki 2002),
and at least for WR~136, the degree of clumping seems to diminish 
in going from the inner wind to the outer wind.

It is not our goal in this study to determine the mass-loss rate; instead,
the unclumped mass-loss rate from Nugis \etal\ is adopted.  To study
the influence of clumping for the IR continuum and line formation,
we assume that the clumping factor is constant over the radii in which
these emission form.  Our model fits to the observations can thus place
constraints on \Dclump.

Also needed for our models is a value of the ``stellar radius'', $R_*$.
For other classes of stars, this would typically be the radius of the
wind base.  For example in the case of Of and OB stars, it is possible
to observe a photospheric radiation field from which one can
derive basic quantities associated with a ``hydrostatic atmosphere'',
such as the stellar rotation speed, surface gravity, and effective
temperature.  Consequently, there is a well-defined radius associated
with the wavelength at which a hot star exhibits peak emission (i.e., in
the UV/EUV).  The only free parameters in the canonical \bb\ velocity
law equation~(\ref{eq:betalaw}) are the value of \bb\ itself and the
terminal velocity \vinf. The latter is generally determined from the
strong and saturated P-Cygni lines in the UV (e.g., Prinja, Barlow, \&
Howarth 1990) or the widths of forbidden lines.

However, the situation is different for WR~stars, fundamentally because
the base of the wind is not observed at any wavelength (Abbott \&
Conti 1987).  The mass-loss rates are so high for the WR~stars that the
winds are optically thick at all wavelengths. This gives rise to the
well-known problem that for such stars, the meaning of stellar radius and
effective temperature are ambiguous (e.g., Castor 1970; Cassinelli 1971).
There have been two approaches to finding values to use for $R_*$ and
$T_{\rm eff}$ in WR~stars. One is that model atmospheres are computed that
have thick winds (e.g., Hamann \& Koesterke 1998ab).  The stellar radius
is chosen to be at the depth at which the Rosseland mean optical depth
reaches a value of 20.  Using this radius and the stellar luminosity, a
stellar temperature \Tstar\ is defined from the Stefan-Boltzmann relation.
The values for WR~136 derived in this way are $\Tstar = 70800$~K and
$\Rstar = 7.5R_\odot$ (Hamann \& Koesterke 1998a).

Nugis \& Lamers (2000) use another approach based on the idea that there
must be a definable hydrostatic star at the base of the wind. These
authors start with an estimate of the radius $R_{\rm evol}$ of WR~stars
provided by stellar evolution modeling.  For example, using formulae
derived by Schaerer \& Maeder (1992) for H-deficient stars, Nugis \&
Lamers find for WR 136, $R_{\rm evol}= 1.2R_\odot$ and $T_{\rm evol}
= 140600~K$.  Their radius is 1/6$^{th}$ as large and their temperature
2 times larger than values derived with the wind modeling approach.
Since WR~136 is not completely H-deficient, the radius
derived from the evolution models is probably underestimated; however,
it should still be smaller than that from the wind-modeling method above.

While the $R_{\rm evol}$ approach may provide a good estimate of the
underlying size of the star, there is a problem in that the optical
depth through the wind to the central star is extremely large, orders
of magnitude greater than unity. Given that the radiation field is not
strongly peaked in the outward direction deep in the wind, we expect
that the rate of increase of the outflow speed is likely to be shallow
just above $R_{\rm evol}$. Hence, we do not consider it likely that this
evolutionary radius is the proper one to use in a $\beta$-law distribution
that describes the conditions in the IR line and continuum formation
regions of the wind.  In other words a velocity law prescription governed
by a single value for $\beta$ is inadequate for describing both the slowly
accelerating inner wind and that in the outer wind.  Nevertheless, it
is convenient to use the $\beta$-law formula to discuss the outer part
of the wind that we can probe with our data. Thus the $b\, R$ product
that appears in the velocity law, would not in this case correspond
to the evolutionary radius of the star, nor would the velocity, $v_0$
necessarily be the sound speed at that radius, but are simply parameters
that complete the description of the outer velocity law. 

To bring closure to our models, we choose to adopt the value
$R_*=5R_\odot$ from the wind modeling of WR~136 by Crowther \& Smith
(1996). Their model is also useful for limiting the possible range
in wind temperature.  For WR~136, Crowther \& Smith (1996) find an
effective temperature of 27700 K at a Rosseland optical depth of 2/3,
but a ``stellar'' temperature of 57700 K at an optical depth of 20
corresponding to a stellar radius of $R_*=5R_\odot$.  In modeling the
forbidden lines at very large radius (100's of $R_*$), an electron
temperature of 13000~K is used (e.g., in Paper~I).  Recent models of
WR~winds show variations in $T_{\rm w}$ with radius, with $T_{\rm w}$
being large at the inner wind, then decreasing with radius to achieve an
asymptotic value (e.g., Dessart \etal\ 2000 and Herald \etal\ 2001).
In this paper we are attempting to extract information from the
ISO observations about the velocity distribution without doing full
radiatively-driven hydrodynamic simulations that might ultimately be
required.  It is likely that the insight developed from this empirical
analysis will provide guidance to more detailed studies.  So as in Hillier
(1987ab), we take the wind to be isothermal in the IR formation region
of interest to us.  We take the temperatures 27700~K and 57700~K as
reasonable limiting values in our models.  The basic stellar and wind
parameters used for WR~136 are summarized in Table~\ref{tab5}.

\subsubsection{The Fit to the Continuum}

For the continuum emission in the vicinity of our line measurements, we
have computed a grid of solutions which are shown in
Figure~\ref{fig9}.  There are two sets of plots:  one for $T_{\rm
w}=27700$ K and one for 57700~K.  The three different curves shown as
solid in the figure are for $\beta=1, 3,$ and $10$ (from left to
right).  The different points are for different clumping factors, with
$\Dclump = 1,3,10,20,$ and $50$, such that $D_c=1$ is
the lowest point on each curve.  The horizontal and vertical dotted
lines demarcate ``allowed zones'' from the observations, representing
the $1\sigma$ confidence range of models permitted by the errors in the
measurements.  For the power-law fits to the continuum slope, we use
the form $f_\lambda = f_0\,(\lambda/\lambda_0)^p$, with $\lambda_0=1$
micron.  As quoted previously, values of $p$  range from $-2.75$ to
about $-2.9$, with $-2.87$ giving the minimum $\chi^2$ fit.  The
absolute flux error for ISO is about 15\%, and propagating errors for
the variation in $p$ results in the allowed range for the flux scale.
Favored models are those that match both the observed continuum slope
(plotted as the abscissa) and the overall flux scale (plotted as the
ordinate), and fall within the indicated error box.  Note
that we do not actually observe the spectrum at 1 $\mu$m; rather the 1
$\mu$m flux level $f_0$ is a back-extrapolated fiducial value based on
the power-law fit to the spectrum.

For comparison we also show models in which the He ionization changes
from \ion{He}{3} to \ion{He}{2} at a radius of $r=30R_*$ (the dashed
line curves). We chose this radius for the change in ionization on the
basis of model calculations by Hillier (1987a) for the well studied
WN~4 star WR~6.  There are probably differences that will arise by the
fact that WR~136 is a WN~6 star instead of WN~4, and because of the fact
that WR~136 is not hydrogen free.  However, we want to investigate the
possible effects of the ionization change and find that Hillier's study
is the best currently available guide.  The location of the ionization
transition requires the detailed models to provide, for example, the
ionizing EUV flux that penetrates through the wind.  We suspect that the
radius of the ionization transition in WR~136 must be relatively large,
because we observe such strong \ion{He}{2} recombination lines.

The effect of an ionization change is to shift the curves of
Figure~\ref{fig9} primarily in a lateral fashion, and more so for the
lower temperature models than for the higher temperature ones.  These new
curves would alter conclusions as to the velocity law. However we note
that rather low clumping factors are still required in the fit shown
in the figure. Clearly, low clumping factors are found both with and
without the ionization transition.  To investigate the other effects
that influence the velocity law diagnostics, we shall assume for the
remainder of the paper that the wind is fully doubly ionized He.

Focusing on the solid lines of Figure~\ref{fig9}, the observed continuum
flux data favor a $\beta$-value of about 5 for the cooler wind, and 3
for the hotter wind. These values are larger than the value $\beta =1$
commonly assumed in WR wind models. However, values as low as $\beta=1$
are not strongly ruled out in Figure \ref{fig9}. The influence of the
temperature can be understood as follows.  A lower temperature reduces
the free-free source function, but increases the free-free opacity.
For a wind that is isothermal and in constant expansion, as applies
to the radio emission, the temperature dependence exactly cancels out
(Cassinelli \& MacGregor 1986).  This cancellation is no longer the
case in the accelerating portion of the wind.  Also, the effect of the
wind acceleration is to modify the density distribution.  In the net
a larger $\beta$ both elevates and steepens the density distribution.
Increasing $\beta$ thus leads to a brighter continuum level with a
steeper power-law slope.

Consider the effect of changing the clumping factor. An increase in
the clumping factor raises the flux level.  However for a {\it fixed}
$\beta$, the enhanced clumping elevates the density such that the
observed continuum emission forms at larger radius and higher speed flow.
Consequently, the continuum slope is actually reduced, since the continuum
forms in a region closer to constant expansion.  Thus to match both the
fiducial flux level and the continuum slope, larger $\beta$-values are
needed for winds of lower temperature.

Note that in the lower temperature case, the power-law slope can
actually fall below the canonical 0.6 value of Wright \& Barlow (1975)
for a constant expansion-velocity wind.  This occurs for two reasons:
(a) both $T_{\rm w}$ and $\lambda$ are sufficiently low that the Planck
function is not entirely in the Rayleigh-Jeans limit, which influences
the slope, and (b) the wavelength dependence of the Gaunt factor in the
IR band is slightly different from that in the radio band.  For the
IR free-free Gaunt factor, we use data from Waters \& Lamers (1984).
These authors computed Gaunt factors down to $10 \microns$ only; however,
we adopt their expressions for the shorter wavelengths of interest with
He$^{++}$ as the dominant contribution (with He mass fraction $Y=8/9$).
We also include free-bound opacity from H$^+$ and He$^{++}$, again using
the Waters \& Lamers expressions.  The free-bound opacity from the IR
and longward has the same scaling as the free-free, and so the effective
Gaunt factor becomes $g_\nu = g_{\rm ff} + b_{\rm fb}$, where $b_{\rm fb}$
is generally much smaller than $g_{\rm ff}$.

\subsubsection{Fits to the Line Profile Data}

To further constrain wind parameters, we have also produced model line
profiles to be compared with the observations.  
To model the lines with the influence of wind acceleration, we again
assume that He$^{++}$ is the dominant ionization state and calculate
the populations of the upper He$^+$ levels by assuming LTE with the
continuum.  The NLTE departure coefficients $b_{\rm k}$ for the upper
levels of interest ($k\ge 7$) are expected to be near unity (e.g.,
Storey \& Hummer 1995).  A grid of line profile calculations are
shown in Figure~\ref{fig10}, with variations in $\beta$, line optical
depth $\tau_l$, and the wind temperature $T_{\rm w}$.  Although these
parameters are not truly independent -- for example, $\tau_l$ will depend
on both $\beta$ and $T_{\rm w}$, they are treated as independent for
the purpose of indicating the range of profile shapes that can result.
The ``canonical'' model to be used as a reference in this figure
(shown as dashed) is for $\beta=1$, $D_c=1$, $T_{\rm w}=30000$~K,
and $\tau_l=1$.  Variation of $D_c$ is not explicitly shown, but its
influence is similar to temperature, in the sense that increasing the
clumping or decreasing the wind temperature both result in overall weaker
emission lines relative to the continuum level.  Note that in our line
modeling, we take the inner wind density to be fixed, with $\rho (R_*) =
\Mdot/4\pi R_*^2 v_0$ a constant.  This means that as $\beta$ is varied,
$b$ also changes to keep $v_0$ fixed, with $v_0= 0.0075\vinf$.

Before applying the line fits to the data, we first demonstrate that
the observed lines are sampling part of the wind acceleration by
overplotting the asymptotic expression for emission line profiles with the
observations in Figure~\ref{fig11} (see Eq.~[\ref{eq:asymp}]).  For our
seven best lines, the emission profile shapes from a wind expanding at
the terminal speed are consistently too broad, and so the line formation
must occur in the lower velocity, accelerating portion of the flow.

In fitting the observed profiles, both $\beta$ and $D_{\rm c}$ are
allowed to vary; however, we must also allow for {\em additional}
broadening to approximate the influence of ``turbulence'' in the wind,
and so a ``broadening parameter'' $\sigma$ is introduced.  This
additonal broadening is motivated by the commonly held fact that the
winds of hot stars are structured.  The velocity distributions of hot
star winds are non-monotonic, with variations associated with wind
shocks that reach outflow speeds that can be significantly greater than
the asymptotic value obtained at large radius.  Line-driven wind
instability effects can produce sharp drops and spikes in the wind
velocity and density distribution in O~star winds (Lucy \& Solomon
1970; Lucy 1982; Owocki, Castor, \& Rybicki 1988; Feldmeier
\etal\ 1997). Similar effects are theoretically expected for WR~winds
(Gayley \& Owocki 1995), and structure is inferred from UV and optical
emission profile variations (e.g., St-Louis \etal\ 1995; L\'{e}pine \&
Moffat 1999).  Of specific relevance to this paper, an analysis of the
\ion{C}{4} doublets near 1550\AA\ reveal blueshifted black troughs and
somewhat larger absorption edge velocities for many classes of hot star
winds, including WR~stars (Prinja \etal\ 1990).  The difference between
the trough and edge velocities, amounting to as much as 25\% of
\vinf\ for WR~winds, is commonly interpreted as arising from the wind
being structured in velocity.  For WR~136, Prinja \etal\ report an edge
velocity of 2700~\kms\ but a terminal speed of 1600~\kms, amounting to
a difference of over 1000~\kms.  Based on our previous comment about
the convolution of Gaussian profile shapes, broadening of this order
would increase the line widths of our model profiles by about 40\%.

To constrain values of $\beta$ and $D_c$ for the IR lines, we have
computed a grid of line profiles to fit our best line, \ion{He}{2}
7--6, which is strong and has excellent S/N.  Using the terminal speed
of \vinf=1490\, \kms\ from Paper~I, we compute the required additional
broadening from $\sigma = \sqrt{HWHM_{\rm obs}^2 -HWHM_{\rm mod}^2}$,
with results shown in Figure~\ref{fig12}, to match the line HW (defined
as the line HW in the red wing to be compared with the observed values
from Tab.~\ref{tab3}).  The solid lines are for the hotter wind, and
the dashed ones for the cooler wind temperature.  In each case, the
upper curve is for $D_c=1$ and the lower for $D_c=3$.  Larger values of
$D_c$ at a fixed $\beta$ will produce broader line widths resulting in
lower $\sigma$-values, but the analysis of the continuum emission
(based on Fig.~\ref{fig9}) rules out $D_c$ greater than about 3.  We
did consider the influence of a transition in the He-ionization at
$r=30R_*$.  With this transition the continuum-normalized line emission
drops by about 10\% or less, and the shape is not dramatically
affected.  The effect is not major because both the line and free-free
opacities are affected so as to be partially compensating when the line
is continuum normalized, and so we continue to assume that the wind is
pure \ion{He}{3}.

In Figure~\ref{fig12}, the lower dotted line is the instrumental
broadening, representing a minimum threshold for $\sigma$.  A strong upper
limit would be 1200~\kms\ based on the UV \ion{C}{4} line absorption
(not shown in the Figure).  Here we have an even better constraint
from the fact that the wind cannot have clumping factors below unity.
At this point, the lines would suggest that $\beta$-values below 1 are
probably not consistent with the observed 7--6 line.

However, we can do better than this.  Wolf-Rayet stars are sources of
X-ray emission (e.g., Pollock 1987; Wessolowski 1996; Ignace, Oskinova,
\& Foullon 2000).  Recent pointed observations of 3 putatively single
WR~stars by XMM-Newton indicate hot gas of at least a few million
Kelvin in the winds (Skinner 2002ab; Ignace, Oskinova, \& Brown 2003).
If arising from wind shocks, as would be the standard explanation,
shock velocity jumps of at least a few hundred \kms\ would be needed to
generate hot gas at these temperatures.  A range of 300-500~\kms\ would
seem to be reasonable, and such variations relative to the mean flow
would lead to ``broadening'' of the IR lines.  The observed X-rays do
not emerge from the same radii as the IR continuum and \ion{He}{2}
lines; the X-rays emerge mainly from farther out in the wind, but only because
of strong photo-absorption by the dense WR~wind.

Consequently, we indicate in Figure~\ref{fig12} with a circled asterisk
a ``best representation'' for the wind of WR~136 that takes into
account (a) low clumping based on the continuum analysis, (b) the
rather broad but finite range of $\beta$ values from the continuum
constraints, and (c) a modest level of $\sigma$ necessary to account for
the fact that WN~stars are X-ray sources.  The result is $D_c=2$, with
$\beta=3$ for the cool wind model and $\beta=2$ for the hot wind model.
Naturally, based on the discussion of the He ionization, and ambiguity
regarding the broadening processes intrinsic to the winds of WR~stars,
$\beta$ is somewhat loosely constrained, although a clumping factor that
is not too large seems required.

Using these preferred values for $\beta$ and $D_c$, Figures \ref{fig13}
and \ref{fig14} show the model profiles (dotted) as overplotted on the
observed emission lines (solid), for the cool and hot wind cases
respectively.  Note that the lines in these figures are continuum
normalized.  A key for the line transitions is provided at bottom
center.  In addition to broadening processes, line blends may influence
some of the observed profiles widths, particularly in the wide
\ion{He}{2} 16--10 line and the wings of \ion{He}{2} 9--7, and the
somewhat blueward broadened \ion{He}{2} 7--6 line.

Inspection of our model results for the lines also confirms the earlier
estimates that the ISO data are not sampling the innermost wind
acceleration.  Optical depth unity in the free-free opacity occurs at a
radius of $v \gtrsim 0.3\vinf$ for our shortest wavelength line.
Still, our analysis places interesting constraints on the value of
$\beta$ and $D_c$ for the extended wind acceleration of WR~136.

\section{DISCUSSION }

We have argued that WR stars should have a significantly different velocity
distribution than OB stars, for which the radiation field is more
forward-peaked near the base of the wind.  The many thick lines and the
small mean free paths deep in the star should cause the initial rise in
velocity to be gradual.  One would expect that a flow which is driven by
multi-line scattering would have a large value of \bb, a conclusion that we
have sought to test using IR~observations of WR~136 obtained with the ISO
satellite. Line and continuum data have been analyzed and are found to be
sensitive primarily to the outer part of the wind at the observed
wavelengths.  Although our ISO data do not probe the deep layers where
multi-line scattering processes are most important, the velocity range
that is probed is still of interest.  If the momentum factor is say 10,
then even in going from 0.7 to 1.0 times the terminal speed, multi-line
scattering is required of the radiation field seeping through the
corresponding wind layers to account for the final acceleration.

We have found that it is possible to obtain a simultaneous fit to both
the continuum and the \ion{He}{2} recombination lines from the ISO data
for WR~136.  We made calculations at two different wind temperatures:
$T_{\rm w}\approx 60000$ K and $T_{\rm w}\approx 30000$~K.  Low clumping
factors with $D_c$ of 1--3 are found.  The wind velocity law is loosely
constrained, but a value of 2--3 appears consistent with the various
constraints imposed by the continuum data, the emission line shapes,
and the expectation of ``turbulent'' broadening in the flow arising from
wind structure.  Our estimate for $\beta \approx$ 2 to 3 falls in the
``gradual acceleration'' range that is perhaps appropriate for multi-line
scattering acceleration.  

For comparison a study of emission line profile variability by L\'{e}pine
\& Moffat (1999) included WR~136 as one of their sources.  They observed
\ion{He}{2} 5411 and draw several interesting conclusions. First, they
find that the turbulence in the wind of WR~136 is typical of other WN
stars in their study.  And second, they derive the product $\beta R_*
\approx 75$.  For the radius of the wind base that we have adopted
($R_*=5 R_\odot$), this would imply a $\beta$ of about 15.  However, as
the authors point out, as well as Koesterke, Hamann, \& Urrutia (2001),
exactly how one relates a $\beta$-value derived from features observed
in Doppler shift space to physical wind velocities is not clear, because
there is almost certainly a flow of wind material past the clumps which
are continually forming and dissipating in the wind.  So although the
data of L\'{e}pine \& Moffat for WR~136 would imply an extensive region
of wind acceleration, the $\beta$ may not describe the ambient wind,
nor does the $R_*$ derived necessarily refer to the base of the wind.
Our method is to infer $\beta$ through the influence of density on the
formation of the IR continuum and \ion{He}{2} lines.  Our results are
perhaps complementary to those of L\'{e}pine \& Moffat, but we are not
studying the same phenomenon.

Our derived low level of clumping is also interesting, since it suggests that
the large scale wind flow that is sampled by the radio observations
and is also claimed to be unclumped (Nugis \etal\ 1998) extends fairly
deep into the wind of WR~136 where the IR emission forms. Perhaps
clumps that exist deep in a wind can be dissipated by wind drag 
with the matter blending into the ambient wind.

In future studies, one could attempt to use somewhat shorter near-IR
spectral data to sample the wind acceleration at smaller radii; however,
this will probably be limited by the fact that high electron optical
depths and strong line-blanketing effectively ``veil'' the innermost
wind acceleration zone.  Perhaps there is no direct technique for
observationally probing the low velocity portion of the densest WR~winds.

\acknowledgements

We express appreciation to the staff at the ISO center. We are also
grateful to Ken Gayley and Henny Lamers for their insightful discussions
on this topic, and an anonymous referee for several helpful comments. The
ISO Spectral Analysis Package (ISAP) is a joint development by the LWS
and SWS Instrument Teams and Data Centers. Contributing institutes are
CESR, IAS, IPAC, MPE, RAL and SRON.  This research was supported by NASA
grants NAG5-3315 and 1207129 (JPL).

\appendix

\section{Analytic Line Profile Shapes for Constant Expansion Velocity Winds}

The emergent intensity $I_\nu$ at impact parameter $p$ within the
line frequencies will be given by the sum of the line component
$I_{\nu,l}$ and the continuum component $I_{\nu,{\rm c}}$.  Adopting
Sobolev theory for the line formation, assuming that the line
optical depth scales with $\rho^2$, and assuming the continuous
emission and opacity is free-free, the respective intensities were
given in eqs.~(\ref{eq:lineinten}) and (\ref{eq:continten}).
Implicit is that the wind is also isothermal.
Further assuming that $S_{\nu,l}=B_\nu$, the total intensity
expression reduces to

\begin{equation}
I_\nu (p) = I_{\nu,l} + I_{\nu,{\rm c}} = B_\nu\,\left[ 1-e^{-(\tau_{\rm max}+
	\tau_S)}\right],
	\label{app:int1}
\end{equation}

\noindent which was also given in eq.~(\ref{eq:simplified}).  For
rays that intercept the stellar core, the emergent stellar
intensity is

\begin{equation}
I_{\nu} (p) = I_{\nu,*} \, e^{-(\tau_{\rm max}+\tau_S)},
	\label{app:int2}
\end{equation}

\noindent where the Sobolev optical depth $\tau_S$ is taken to be
zero for red-shifted frequencies in the line, since those resonance
points all lie aft of the star.

In the case of constant wind expansion, the line profile shape
including the influence of the continuous opacity is analytic in a
certain limit as we now show.  First, the Sobolev optical depth is given
by

\begin{equation}
\tau_S = \tau_l\,x^{-4}\,w^{-2}\,\left[\mu^2\,\frac{dw}{dx} +
	(1-\mu^2)\,\frac{w}{x} \right]^{-1}.
\end{equation}

\noindent For constant expansion-velocity, this reduces to

\begin{equation}
\tau_S = \tau_l\,x^{-3}\,(1-\mu^2) = \tau_l\,\sin\theta\,p^{-3}.
\end{equation}

\noindent where we have used the fact that $p=x\sin \theta$.
The free-free optical depth is given by

\begin{equation}
\tau_{\rm ff} = \tau_{\rm c}\,\int_z^\infty\,\frac{dz}{x^4\,w^2}.
\end{equation}

\noindent For constant expansion-velocity, and re-expressing the
integral in terms of $p$ and $\theta$, an analytic solution exists
giving analytically to yield

\begin{equation}
\tau_{\rm ff} = \frac{1}{2}\,\tau_{\rm
c}\,p^{-3}\,(\theta-\sin\theta\,\cos\theta).
\end{equation}

\noindent The value of $\tau_{\rm max}$ in eqs.~(\ref{app:int1}) and
(\ref{app:int2}) depends on $p$ as follows:

\begin{equation}
\tau_{\rm max} = \cases{ \frac{\tau_{\rm c}}{2p^3}\,(\theta-\sin\theta\,\cos\theta) & ($p < 1$), \cr
	\frac{\pi}{2}\,\frac{\tau_{\rm c}}{p^3} & ($p\ge 1$)  }
\end{equation}

The observed flux of line emission at any frequency in the line will be

\begin{eqnarray}
f_\nu & = & 2\pi\,\frac{R^2}{D^2}\,\int_{w_{\rm z}} \, I_\nu(p)\,p\,dp \\
      & = & 2\pi\,B_\nu(T_w)\,\frac{R^2}{D^2}\,\left\{ \int_0^{p_{\rm min}} \,
		\frac{I_{\nu,*}(T_*)}{B_\nu(T_w)}\,p\,dp\,\left[1-\exp(-0.5
		\tau_{\rm c}\,h(\theta)\,p^{-3}) \right] \right. \nonumber \\ 
      &   & + \int_{p_{\rm min}}^1 \, \frac{I_{\nu,*}(T_*)}{B_\nu(T_w)}\,p\,dp\,\left[1-\exp(-0.5\tau_{\rm c}\,h(\theta)
		\,p^{-3}-\tau_S\,\sin\theta\,p^{-3}) \right] \nonumber\\ 
      &   & \left. + \int_1^\infty \, p\,dp\,\left[1-\exp(-0.5\pi\,\tau_{\rm c}\,
		\,p^{-3}-\tau_S\,\sin\theta\,p^{-3}) \right] \right\}, 
	\label{app:obsflux} 
\end{eqnarray}

\noindent where

\begin{equation}
h(\theta) = \theta - \sin\theta\cos\theta.
\end{equation}

\noindent For blue-shifted frequencies, $p_{\rm min} = \sin \theta$
since the isovelocity surfaces are cones of constant opening angle that
intercept a spherical core photosphere.  Although all of the stellar
photosphere suffers absorption via the free-free opacity, only the
annulus of the star from $\theta$ to $\pi/2$ will be diminished by the
line opacity.  At red-shifted frequencies, $p_{\rm min} = 1$ and the
second integral in (\ref{app:obsflux}) vanishes; there is no line emission
for $p<1$ owing to stellar occultation.

To solve these integrals, we make the change of variable $u^3 =
\tau_{\rm c}/p^3$.  The integrals become

\begin{eqnarray}
f_\nu & = & 2\pi\,\tau_{\rm c}^{2/3}\,B_\nu(T_w)\,\frac{R^2}{D^2}\,\left\{ \int_{\tau_{\rm c}^{1/3}p_{\rm min}^{-1}}^\infty \,
		u^{-3}\,du\,\frac{I_{\nu,*}(T_*)}{B_\nu(T_w)}\,\left[1-\exp(-0.5
		h(\theta)\,u^3) \right] \right. \\ \nonumber
      &   & \int_{\tau_{\rm c}^{1/3}}^{\tau_{\rm c}^{1/3}/p_{\rm min}} \, u^{-3}\,du\,\frac{I_{\nu,*}(T_*)}{B_\nu(T_w)}\,\left[1-\exp(-0.5h(\theta)
		\,u^3-\tau_S\,\tau_{\rm c}^{-1}\sin\theta\,u^3) \right] \\ \nonumber
      &   & \left. \int_0^{\tau_{\rm c}^{1/3}} \, u^{-3}\,du\,\left[1-\exp(-0.5\pi
		\,u^3-\tau_S\,\tau_{\rm c}^{-1}\sin\theta\,u^3) \right] \right\} \\
\end{eqnarray}

\noindent It is important to note that the frequency information
is contained in the functions of $\theta$ via $w_{\rm z} = - \mu$,
and also that these functions are constants of the integration
which is over $u$.

There is in fact an analytic solution to this form of integral
given in Gradshteyn \& Ryzhik (1994; Eq.~3.478, \#2), for which
the relevant result is

\begin{equation}
\int_0^\infty\,u^{-3}\,du\,\left(1-e^{-au^3}\right) = 1.5\,\Gamma(4/3)\,a^{2/3}.
\end{equation}

\noindent This solution is applicable to the case at hand
in the limit that $\tau_{\rm c}^{1/3} \gg 1$,
with the result

\begin{equation}
f_\nu = 3\Gamma(4/3)\,\pi\,B_\nu(T_w)\,\tau_{\rm c}^{2/3}\,\left(\frac{\pi}{2}
	+\frac{\tau_l}{\tau_{\rm c}}\,\sqrt{1-w_{\rm z}^2}\right)^{2/3}.
\end{equation}

\noindent Interestingly, for a strong line with $\tau_l \gg \tau_{\rm c}$,
the profile shape reduces to

\begin{equation}
f_\nu = 3\Gamma(4/3)\,\pi\,B_\nu(T_w)\,\tau_l^{2/3}\,(1-w_{\rm z}^2)^{1/3}.
	\label{eq:asymp}
\end{equation}

\begin{figure}
\plotone{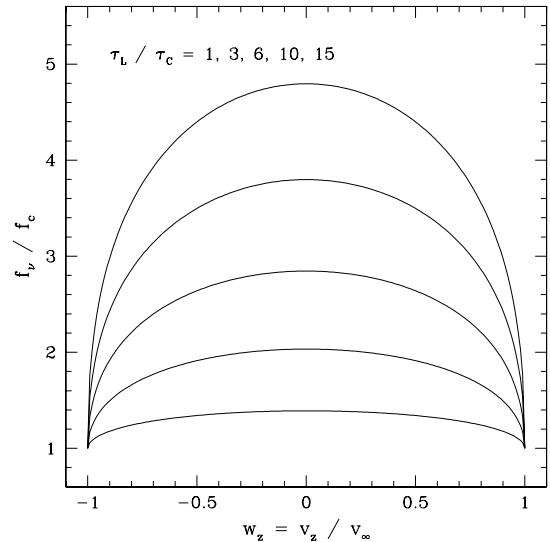}
\caption{A sequence of theoretical emission profiles for 
a constant expansion-velocity wind as the parameter $\tau_l/\tau_{\rm c}$ is
varied, where $\tau_l$ is the scale factor for the line optical depth,
and $\tau_{\rm c}$ is the scale factor for the free-free optical depth. 
The profiles are continuum normalized and plotted in velocity units 
normalized to the wind terminal speed.}
\label{fig4}
\end{figure}

\begin{figure}
\plotone{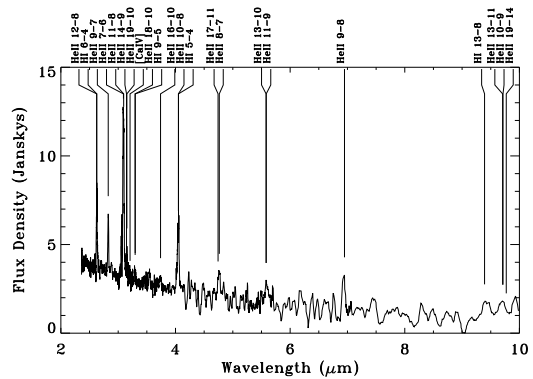}
\caption{The de-reddened low resolution (SWS01) spectrum of WR 136.
Major emission features, including the \ion{He}{2} recombination
lines used in the analysis in this paper are identified. Beyond
about 6 \microns, the signal-to-noise ratio is seriously degraded,
precluding derivation of the velocity law in the outermost regions
of the stellar wind.} \label{fig5} 
\end{figure}

\begin{figure}
\plotone{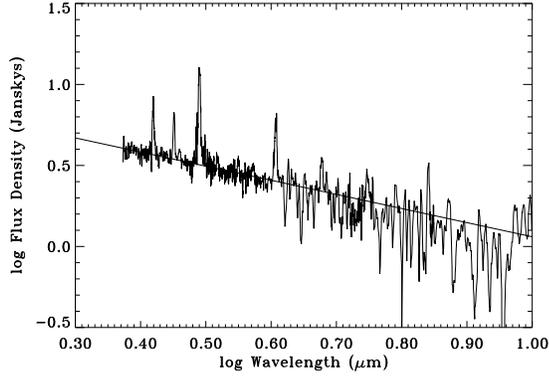}
\caption{The de-reddened continuum flux distribution. This continuum
is fit with a power law to give the index,
$\gamaF = d \log f_\lambda/d \log \lam = - 2.87$.}
\label{fig6}
\end{figure}

\begin{figure}
\plotone{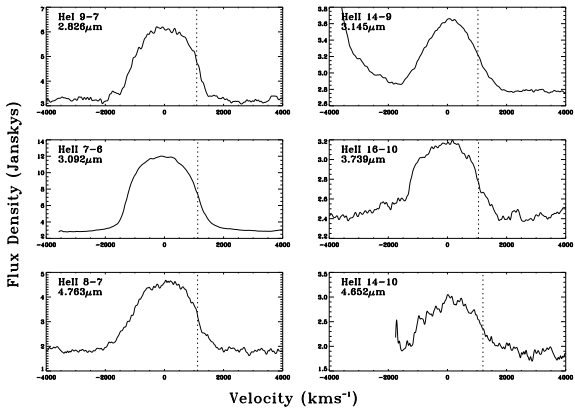}
\caption{High resolution (SWS06) spectra of six \ion{He}{2}
recombination lines. The spectra have not been de-reddened.  With
reference to the red wing, the HWHM velocities measured for our
analysis are indicated by vertical dashed lines.  } \label{fig7}

\end{figure}

\begin{figure}
\plotone{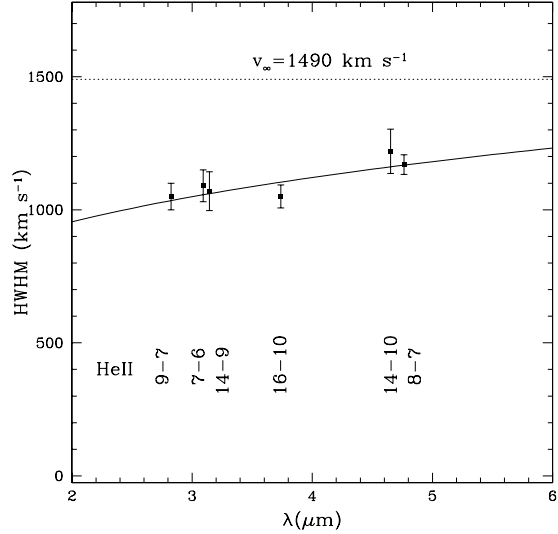}
\caption{Velocity as a function of wavelength as determined by the
HWHM of the \ion{He}{2} recombination lines. The terminal speed
$\vinf = 1490~\kms$ is indicated as a dotted horizontal line. The
distribution is fit with a power law to give the index, $\gamav =
d \log {\rm HWHM}/d \log \lam= 0.22$.}

\label{fig8}
\end{figure}

\begin{figure}
\plottwo{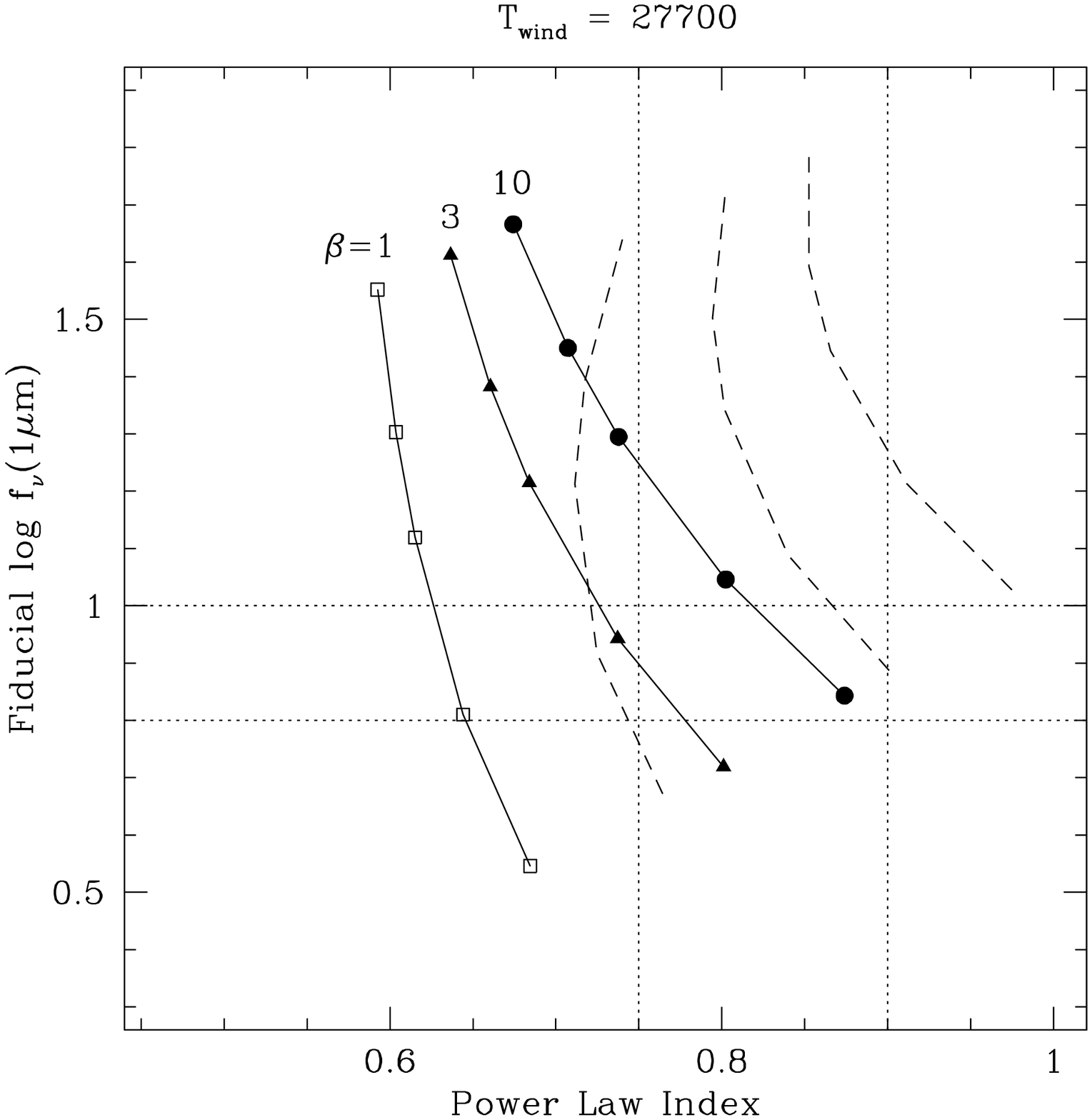}{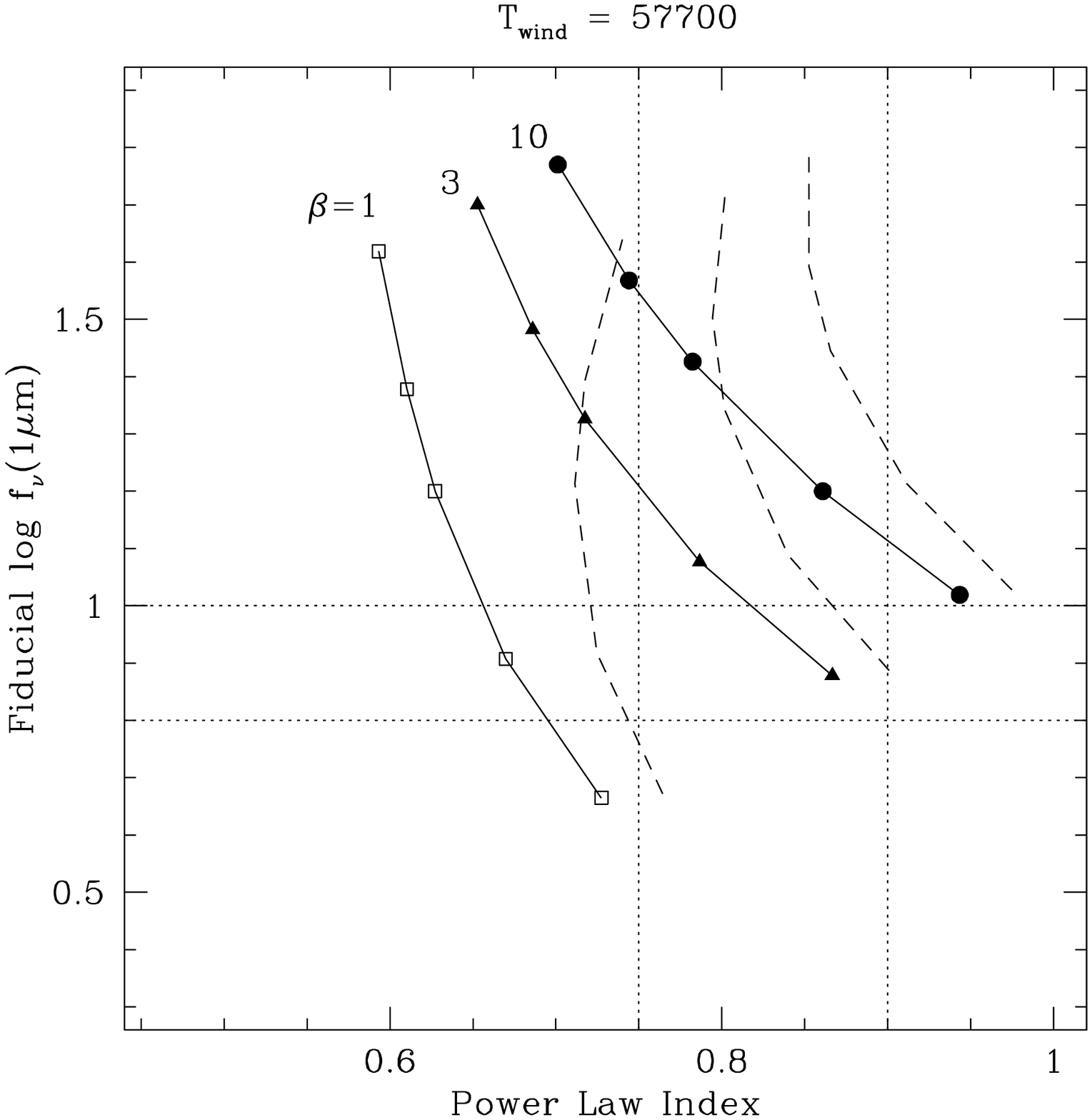}
\caption{
Shown as dotted lines are the $1\sigma$ ``allowed'' zones from the ISO
data.  Model calculations for the continuum slope (abscissa) and flux
level (ordinate).  Left is for a modest wind temperature of 27700~K and
right for 57700~K.  The three solid curves in each panel are for $\beta=1$,
3, and 10.  The points are for different clumping
factors $D_c = 1,3,10,20,$ and 50, moving from bottom to top.  The dashed
lines are for models with a transition from \ion{He}{3} to
\ion{He}{2} at $r=30R_*$.  Although the curves can shift significantly
(especially for lower temperature), the shift is mostly lateral such that
low values of $D_c$ are favored with or without the Helium ionization
transition.  }

\label{fig9}
\end{figure}

\begin{figure}
\plotone{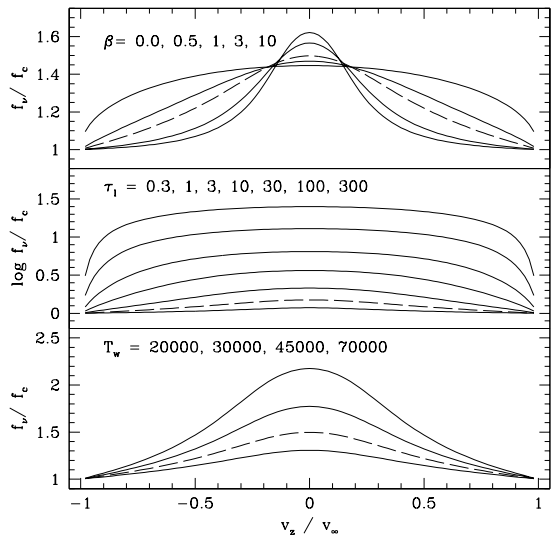}
\caption{A sequence of model emission line profiles to illustrate
the effects of varying different parameters.  At top, line profile
calculations with higher $\beta$ have narrower half-widths.  At middle,
line profiles of higher optical depth $\tau_l$ have stronger emission.
At bottom, winds of higher temperature $T_{\rm w}$ yield relatively
stronger lines.  (The same would apply for winds of lower $D_c$-values.)
The lines are continuum normalized, so one should be aware that
the variation of $\beta$ and $T_{\rm w}$ changes the continuum.
Even though these parameters are not in fact independent, by varying
them independently, we obtain a grid of models to illustrate the range
of line widths and strengths that can result.}

\label{fig10}
\end{figure}

\begin{figure}
\plotone{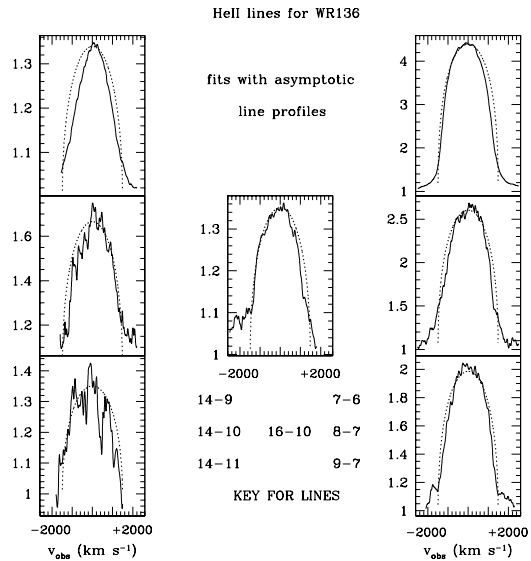}
\caption{Application of the constant expansion-velocity line profile shape
to the observed emission lines (plotted as continuum normalized).
These theoretical asymptotic forms actually match the observed
lines fairly well, although typically being somewhat broader
(especially clear in the 14-11 line).  Even so, better fits arise
when a velocity law is included, especially in light of the fact
that there is some blending that likely adds extra broadening to
the lines of interest. } \label{fig11}

\end{figure}

\begin{figure}
\plotone{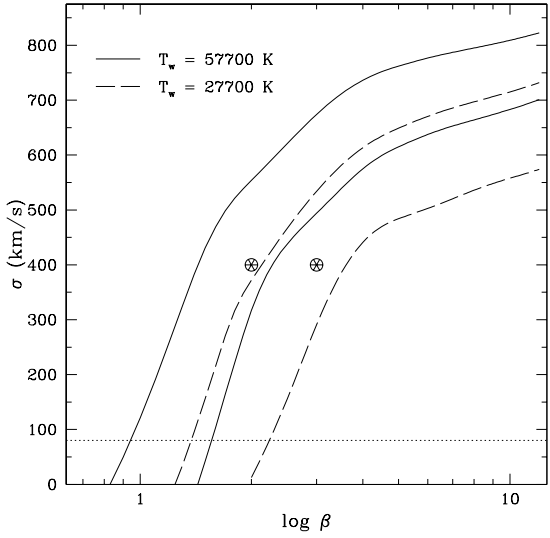}
\caption{A plot of the line broadening $\sigma$ required to bring model
line widths into agreement with that observed in the \ion{He}{2}
7--6 line.  The solid lines are for $T_{\rm w}=57700$~K and the dashed
ones for $T_{\rm w}=27700$~K.  In each case the clumping factor $D_c=1$
for the upper curve and $D_c=3$ for the lower curve.  The lower line
is the minimum $\sigma$ set by the instrumental broadening.  The two
starred points indicate models with $\beta$, $D_c$, and $\sigma$ that meet all
the requirements fo the data, with the point at left for the higher
temperature wind.
} \label{fig12}

\end{figure}

\begin{figure}
\plotone{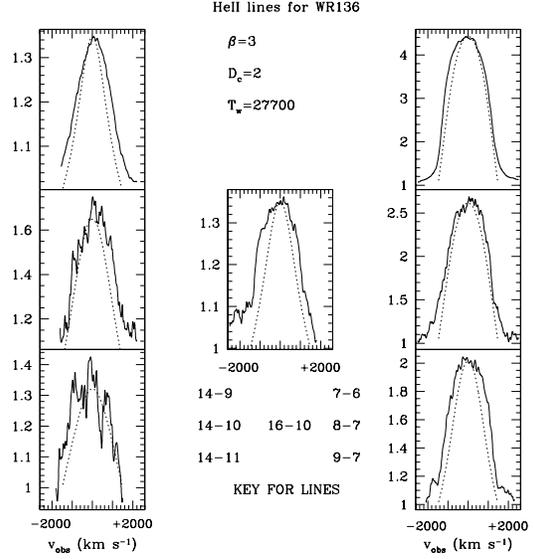}
\caption{Shown as dashed lines are the pre-broadened line models
with $\beta=3$, $D_c=2$, and a wind
temperature of $T_{\rm w}=27700$~K overplotted with the observed
\ion{He}{2} lines. } \label{fig13} 
\end{figure}

\begin{figure}
\plotone{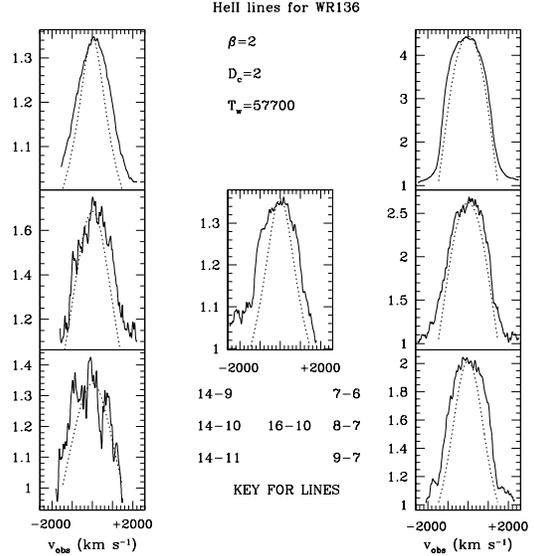}
\caption{As in Fig.~\ref{fig13}, but for models with 
$\beta=2$, $D_c=2$, and a wind
temperature of $T_{\rm w}=57700$~K. }
\label{fig14} 
\end{figure}

\noindent The results of this derivation show that symmetric emission
profiles result, which are overall similar in appearance to those
observed for WR~136.  The rounded shape of the profiles implies that a
measurement of the widths of these lines would seriously underestimate
the speed of the wind flow.  Examples of the normalized profile shape of
eq.~(\ref{eq:asymp}) for various values of the ratio $\tau_l/\tau_{\rm
c}$ are given in Figure~\ref{fig4}.

\onecolumn

\clearpage

\begin{deluxetable}{lrrrrr}
\tabletypesize{\normalsize}
\tablecaption{Journal of Observations\label{tab1}}
%\vspace{.5cm}
\tablewidth{0pt}
\tablehead{
\colhead{Observation}   & \colhead{Spectrograph}  &
\colhead{Date} &\colhead{Observation}  & \colhead{Duration } \\
\colhead{Name}   & \colhead{}  &
\colhead{} &\colhead{Number}  & \colhead{(seconds)} } 
\startdata
WR 136      &SWS06 &1 Dec 1996 &38101710 &12872 \\
WR 136      &SWS01 &1 Dec 1996 &38101711 &1140 \\
WR 136 Off  &SWS01 &1 Dec 1996 &38101712 &1140 
\enddata

\end{deluxetable}

\begin{table}
\begin{center}
\caption{Line Identifications\label{tab2}}
\end{center}
\begin{tabular}{lllll}
\tableline\tableline
$\lambda$ & Principal Line & Blends at $\Delta v$ (\kms)&  Notes\\
\tableline

 2.83 & \ion{He}{2} 9--7 & \ion{C}{4} 15--9 at $+750$ & features near the blue and red \\
     & 	   &			    &   wings are possibly weak \ion{He}{1} and \\
     & 	   &			    &  \ion{He}{2} lines\\
 & & & \\

 3.09 &	\ion{He}{2} 7--6 & \ion{N}{3} 8--7 at $-390$ &	the N line could be a significant \\ 
     & 	   & \ion{C}{3} 9--8 at $-290$ & contributor to the line flux \\
     &	   & \ion{C}{4} 14--12 at 0    &  \\    
 & & & \\

     &     &   \ion{He}{2} 11--8 at $+350$ &  since \ion{He}{2} 14--9 have emission\\

     & 	   &		   & at about 30\% of $f_{cont}$, this feature \\
     & 	   &		   & likely has a slight influence on \\
     & 	   &		   & the width of \ion{He}{2} 7--6 \\
     &     &     & \\

 3.15 &	\ion{He}{2} 14--9 & \ion{He}{2} 19--10 at $+560$ &  blend may account for difficulty in \\
     & 	   &			   & matching the red wing width \\
 & & & \\

 3.74 &	\ion{He}{2} 16--10 & \ion{H}{1} (Pf $\gamma$) at 0 &                         \\

&     & \ion{He}{1} 8--5 at 0  & a sequence of transitions between \\
     & 	   &			    & different terms in the 8--5 line spread \\
     & 	   &			    & from $-1000$ to 0, which may account\\
     & 	   &			    & for the slight blue wing broadening\\
 & & & \\

 4.65 &	\ion{He}{2} 14--10 & \ion{H}{1} (Pf $\beta$) at 0 &                 \\
     &	     &	\ion{He}{2} 22--12 at $+1225$ &               \\
 & & & \\

 4.76 &	\ion{He}{2} 8--7 & \ion{He}{2} 17--11 at $-1350$ &                \\
 & & & \\

 7.21 &	\ion{He}{2} 14--11\tablenotemark{(a)} &	\ion{He}{2} 19--13 at $+1460$ &     \\

\tableline

\tablenotetext{a}{We include 14--11 so that there are 3 lines with
upper level 14.  The 14--9 and 14--11 do not blend with H, but
14--10 does.  In comparing the two with odd lower level, we can
confirm whether the lines are thin. By comparing to the one with
even lower level, we can infer the H/He ratio.}

\end{tabular}
\end{table}

\begin{table}
\begin{center}
\caption{\HeII\ Recombination Line Data for WR 136\label{tab3}}
\end{center}
\begin{tabular}{crccccc}
\tableline\tableline
Wavelength  & Transition  & HWHM$^a$ & Continuum Flux & EW \\ %& $gf$ & $\lambda\, EW/gf$ \\
(\microns)  & (\ion{He}{2}) & (\kms)   & (Jy)           & (\AA) \\ %&  & \\
\tableline
2.826 & 9--7   & 1050 & 3.44 & 210 \\ %& 25.6 & 23 \\
3.092 & 7--6   & 1090 & 3.20 & 790 \\ %& 102  & 24 \\
3.145 & 14--9  & 1070 & 3.17 & 71  \\ %& 4.73 & 47 \\
3.739 & 16--10 & 1050 & 2.79 & 97  \\ %& 3.91 & 93 \\
4.652 & 14--10 & 1220 & 2.37 & 220 \\ %& 11.0 & 94 \\
4.764 & 8--7   & 1170 & 2.32 & 570 \\ %& 158  & 17 \\
7.208$^b$ & 14--11 & ---  & --- & 180 \\ %& 30   & 43 \\
\tableline
\end{tabular}

$^a$ The HWHM is evaluated for the red wing.

$^b$ The \ion{He}{2} 14--11 line and neighboring continuum are not sufficiently
reliable to quote a line width or continuum flux value, although it is shown 
for the model line fits described in \S 3.3.4.

\end{table}

\begin{table}
\begin{center}
\caption{Mass-Loss Rate Estimates for WR 136\label{tab4}}
\end{center}
\begin{tabular}{lccc}
\tableline\tableline
 Reference   & \Mdot               & Waveband & Flow Type \\
             & ($10^{-5}$\Msunyr)  &             &           \\
\tableline
 Abbott \etal\ (1986)        & 3.2  & Radio free-free & smooth \\
 Barlow \& Cohen (1975)      & 2.3  & Radio free-free & smooth \\
 Crowther \& Smith (1996)    & 12.5 & NIR lines       & smooth \\
 Hamann \& Koesterke (1998)  & 12.5 & Optical/UV lines & smooth \\
 Nugis \etal\ (1998)         & 6.25 & Radio           & clumped$^a$ \\
\tableline
\end{tabular}

{\small $^a$ These authors allow for clumping in their analysis, but conclude
that the outer wind is unclumped.}

\end{table}

\begin{table}
\begin{center}
\caption{Adopted Parameters for WR 136\label{tab5}}
\end{center}
\begin{tabular}{lccc}
\tableline\tableline
Parameter & Value \\
\tableline
Subtype$^a$ & WN6 \\
$d^b$ & 1.82 kpc \\
$T_*^c$ & 57700~K \\
$R_*^c$ & $5R_\odot$ \\
$L_*$ & $2.5\times 10^5 L_\odot$\\
$\dot{M}^d$ & $6.25\times 10^{-5} M_\odot\;{\rm yr^{-1}}$ \\
$v_\infty^e$ & 1490 \kms\ \\
$T_{\rm w}$ & 27,700--57,700 K \\
$X^f$ & 0.12 \\
\tableline
\end{tabular}

{\small

$^a$ From van der Hucht (2001). 

$^b$ From Lundstr\"{o}m \& Stenholm (1984.

$^c$ From Crowther \& Smith (1996) that are assumed for the base
of the wind.

$^d$ From Nugis \etal\ (1998).

$^e$ From Paper~I.

$^f$ $X$ is the hydrogen mass-fraction.

}

\end{table}

\end{document}